\documentclass[11pt,onecolumn,onecolumn]{IEEEtran}
\usepackage[LGR,T1]{fontenc}
\usepackage[latin9]{inputenc}
\usepackage{amsmath}
\usepackage{amsthm}
\usepackage{amssymb}
\usepackage{graphicx}

\makeatletter
\theoremstyle{plain}
\newtheorem{thm}{\protect\theoremname}
\theoremstyle{plain}
\newtheorem{conjecture}{\protect\conjecturename}
\theoremstyle{plain}
\newtheorem{prop}{\protect\propositionname}
\theoremstyle{remark}
\newtheorem{claim}{\protect\claimname}
\theoremstyle{plain}
\newtheorem{lem}{\protect\lemmaname}

\usepackage{color}

\usepackage{cite}

\usepackage[mathscr]{eucal}
\usepackage{epsfig,epsf,psfrag}
\usepackage{amssymb,amsmath,amsthm,latexsym}
\usepackage{amsmath,graphicx,xcolor,url}
\usepackage[caption=false]{subfig} 
\usepackage{fixltx2e}
\usepackage{array}
\usepackage{verbatim}
\usepackage{bm}
\usepackage{algorithmic}
\usepackage{algorithm}
\usepackage{verbatim}
\usepackage{textcomp}
\usepackage{mathrsfs,overpic}
\usepackage{epstopdf}
\usepackage{amsfonts}
\usepackage[numbers]{natbib}

\usepackage{tikz}
\usepackage{float}
\usepackage{tabularx}
\usepackage{multirow}
\usetikzlibrary{patterns}

\newcommand{\Unif}{\mathrm{Unif}}

\def\1{\mathbf{1}}

\catcode`~=11 \def\UrlSpecials{\do\~{\kern -.15em\lower .7ex\hbox{~}\kern .04em}} \catcode`~=13 

\allowdisplaybreaks[1]

\newcommand{\calM}{\mathcal{M}}

\newcommand{\bM}{\mathbf{M}}

\newcommand{\bU}{\mathbf{U}}

\newcommand{\bV}{\mathbf{V}}

\newcommand{\bbR}{\mathbb{R}}

\DeclareMathAlphabet{\mathbsf}{OT1}{cmss}{bx}{n}
\DeclareMathAlphabet{\mathssf}{OT1}{cmss}{m}{sl}

\DeclareSymbolFont{bsfletters}{OT1}{cmss}{bx}{n}  
\DeclareSymbolFont{ssfletters}{OT1}{cmss}{m}{n}
\DeclareMathSymbol{\bsfGamma}{0}{bsfletters}{'000}
\DeclareMathSymbol{\ssfGamma}{0}{ssfletters}{'000}
\DeclareMathSymbol{\bsfDelta}{0}{bsfletters}{'001}
\DeclareMathSymbol{\ssfDelta}{0}{ssfletters}{'001}
\DeclareMathSymbol{\bsfTheta}{0}{bsfletters}{'002}
\DeclareMathSymbol{\ssfTheta}{0}{ssfletters}{'002}
\DeclareMathSymbol{\bsfLambda}{0}{bsfletters}{'003}
\DeclareMathSymbol{\ssfLambda}{0}{ssfletters}{'003}
\DeclareMathSymbol{\bsfXi}{0}{bsfletters}{'004}
\DeclareMathSymbol{\ssfXi}{0}{ssfletters}{'004}
\DeclareMathSymbol{\bsfPi}{0}{bsfletters}{'005}
\DeclareMathSymbol{\ssfPi}{0}{ssfletters}{'005}
\DeclareMathSymbol{\bsfSigma}{0}{bsfletters}{'006}
\DeclareMathSymbol{\ssfSigma}{0}{ssfletters}{'006}
\DeclareMathSymbol{\bsfUpsilon}{0}{bsfletters}{'007}
\DeclareMathSymbol{\ssfUpsilon}{0}{ssfletters}{'007}
\DeclareMathSymbol{\bsfPhi}{0}{bsfletters}{'010}
\DeclareMathSymbol{\ssfPhi}{0}{ssfletters}{'010}
\DeclareMathSymbol{\bsfPsi}{0}{bsfletters}{'011}
\DeclareMathSymbol{\ssfPsi}{0}{ssfletters}{'011}
\DeclareMathSymbol{\bsfOmega}{0}{bsfletters}{'012}
\DeclareMathSymbol{\ssfOmega}{0}{ssfletters}{'012}

\newcommand{\Bern}{\mathrm{Bern}}

\DeclareMathOperator{\rank}{rank}

\newcommand{\qednew}{\nobreak \ifvmode \relax \else
      \ifdim\lastskip<1.5em \hskip-\lastskip
      \hskip1.5em plus0em minus0.5em \fi \nobreak
      \vrule height0.75em width0.5em depth0.25em\fi}

\usepackage{bm,bbm}

\newcommand{\bone}{\mathbbm{1}}

\allowdisplaybreaks[1]
\providecommand{\claimname}{Claim}
\providecommand{\conjecturename}{Conjecture}
\providecommand{\lemmaname}{Lemma}
\providecommand{\propositionname}{Proposition}
\providecommand{\theoremname}{Theorem}

\makeatother

\providecommand{\claimname}{Claim}
\providecommand{\conjecturename}{Conjecture}
\providecommand{\lemmaname}{Lemma}
\providecommand{\propositionname}{Proposition}
\providecommand{\theoremname}{Theorem}

\begin{document}
\title{Rényi Common Information for Doubly Symmetric Binary Sources }
\author{Lei Yu, \IEEEmembership{Member,~IEEE}\thanks{L. Yu is with the School of Statistics and Data Science, LPMC, KLMDASR,
and LEBPS, Nankai University, Tianjin 300071, China (e-mail: leiyu@nankai.edu.cn).
This work was supported by the National Key Research and Development
Program of China under grant 2023YFA1009604, the NSFC under grant
62101286, and the Fundamental Research Funds for the Central Universities
of China (Nankai University) under grant 054-63243076.}}

\maketitle
\begin{abstract}
In this note, we provide analytic expressions for the Rényi common
information of orders in $(1,\infty)$ for the doubly symmetric binary
source (DSBS). Until now, analytic expressions for the Rényi common
information of all orders in $[0,\infty]$ have been completely known
for this source. We also consider the Rényi common information of
all orders in $[-\infty,0)$ and evaluate it for the DSBS. We provide
a sufficient condition under which the Rényi common information of
such orders coincides with Wyner's common information for the DSBS.
Based on numerical analysis, we conjecture that there is a certain
phase transition as the crossover probability increasing for the Rényi
common information of negative orders for the DSBS. Our proofs are
based on a lemma on splitting of the entropy 
and the analytic expression of relaxed Wyner's common information. 
\end{abstract}

\section{Introduction}

The common information problem concerns determining the amount of
common randomness required to simulate two correlated sources in a
distributed fashion. The KL-approximate version of such a problem
was first studied by Wyner \cite{WynerCI}, who used the normalized
relative entropy (Kullback-Leibler (KL) divergence) to measure the
approximation level (discrepancy) between the simulated joint distribution
and the joint distribution of the original correlated sources. The
present author and Tan \cite{YuTan2018,yu2020corrections} generalized
Wyner's result such that the approximation level is measured in terms
of the Rényi divergence, thus introducing the notion of Rényi common
information. Kumar, Li, and El Gamal \cite{KLE2014} proposed a quantity,
called exact common information, for which the generated source is
required to follow the desired distribution exactly, and meanwhile,
the common randomness is allowed to be compressed by variable-length
codes. In \cite{YuTan2020_exact}, by using the equivalence between
Rényi common information of order $\infty$ and the exact common information,
the present author and Tan completely characterized the exact common
information (or equivalently, the Rényi common information of order
$\infty$) for doubly symmetric binary sources (DSBSes), and showed
that for this source, the exact common information is strictly larger
than Wyner's common information. In fact, until now, DSBSes are the
unique class of sources for which the exact common information is
known to be strictly larger than Wyner's common information.

The papers \cite{YuTan2018,yu2020corrections} mainly focused on the
Rényi common information of orders in $[0,2]$, and the paper \cite{YuTan2020_exact}
only focus on the Rényi common information of order $\infty$. For
DSBSes, the analytic expression for the Wyner common information (i.e.,
Rényi common information of order $1$) was given in \cite{WynerCI},
the analytic expressions for the Rényi common information of orders
in $[0,1)$ was given in \cite{YuTan2018,yu2020corrections}, and
the analytic expression for the Rényi common information of order
$\infty$ was given in \cite{YuTan2020_exact}. The analytic expressions
for the Rényi common information of other orders (i.e., orders in
$[-\infty,0)\cup(1,\infty)$) were previously unknown.

\subsection{Our Contributions}

In this note, we provide analytic expressions for the Rényi common
information of orders in $(1,\infty)$ for DSBSes. That is, until
now, analytic expressions for the Rényi common information of all
orders in $[0,\infty]$ have been completely known for this source. 

We also consider the Rényi common information of all orders in $[-\infty,0)$
and evaluate it for the DSBS. We provide a sufficient condition under
which the Rényi common information of such orders coincides with Wyner's
common information for the DSBS. This sufficient condition is numerically
verified to be true for $\epsilon\ge\epsilon_{0}$, where $\epsilon_{0}\approx0.05510465170298144$.

According to numerical results, we conjecture that there is a certain
phase transition for the Rényi common information of negative orders
for the DSBS. Specifically, we conjecture that there is a certain
phase transition occurring around $\epsilon=\epsilon_{0}$: For $\epsilon\ge\epsilon_{0}$,
the $-\infty$-Rényi common information is equal to the Wyner common
information; while for $\epsilon<\epsilon_{0}$, the $-\infty$-Rényi
common information is strictly larger than the Wyner common information.
This conjecture is interesting if it is ture, since this kind of phase
transitions is uncommon in information theory. 

\subsection{Notations}

Let $\mathcal{X},\mathcal{Y}$ be finite alphabets. We use $\pi_{X},P_{X},Q_{X}$
to denote distributions on $\mathcal{X}$. Let $\mathcal{P}(\mathcal{X})$
be the set of distributions on $\mathcal{X}$. For two distributions
$P,Q$ on the same space, the {\em Rényi divergence of order $\alpha\in(-\infty,0)\cup(0,1)\cup(1,\infty)$}
is defined as 
\begin{align}
D_{\alpha}(P\|Q) & :=\frac{1}{\alpha-1}\log\sum_{x}P(x)^{\alpha}Q(x)^{1-\alpha}.\label{eq:-40}
\end{align}
Throughout, the convention $0\cdot\infty=0$ is adopted and $\log$
is to the base $2$. The Rényi divergence of order $\alpha\in\{0,1,\pm\infty\}$
is defined by continuous extension. In particular, 
\[
\lim_{\alpha\to1}D_{\alpha}(P\|Q)=D(P\|Q),
\]
where $D(P\|Q)$ is the {\em relative entropy} given by 
\begin{equation}
D(P\|Q):=\sum_{x\in\mathrm{supp}(P)}P(x)\log\frac{P(x)}{Q(x)}.\label{eq:-19-1}
\end{equation}
Denote $H(X)$ or $H(P)$ as the entropy of $X\sim P$. Denote $I(X;Y)=H(X)+H(Y)-H(XY)$
as the mutual information between $X$ and $Y$. Denote $\bar{a}:=1-a$.
When there is no ambiguity, we denote the binary entropy function
as $H(a):=-a\log a-\bar{a}\log\bar{a}$ for $a\in[0,1]$. Denote the
binary relative entropy function as $D(a\|b):=a\log\frac{a}{b}+\bar{a}\log\frac{\bar{a}}{\bar{b}}$.
Denote $a*b=a\bar{b}+\bar{a}b$ as the binary convolution of $a$
and $b$.

Let $\pi_{XY}$ be a distribution on finite alphabet $\mathcal{X}\times\mathcal{Y}$.
Let $\pi_{XY}^{\otimes n}$ be the $n$-fold product of $\pi_{XY}$
with itself. Given $\pi_{XY}$ and $\alpha\in[-\infty,\infty]$, the
Rényi common information of order $\alpha$ \cite{YuTan2018,yu2020corrections},
denoted as $T_{\alpha}(\pi_{XY})$ (resp. $\widetilde{T}_{\alpha}(\pi_{XY})$),
is defined as the minimum rate needed to ensure that the Rényi divergence
$D_{\alpha}(P_{X^{n}Y^{n}}\|\pi_{XY}^{\otimes n})$ (resp. $\frac{1}{n}D_{\alpha}(P_{X^{n}Y^{n}}\|\pi_{XY}^{\otimes n})$)
vanishes asymptotically as $n\to\infty$. The case of $\alpha=1$
corresponds to Wyner's common information, which was proven by Wyner
\cite{WynerCI} to be equal to 
\begin{equation}
C_{\mathsf{W}}(\pi_{XY})=\min_{P_{XYW}:\,P_{XY}=\pi_{XY},\,X\leftrightarrow W\leftrightarrow Y}I(XY;W),\label{eqn:CWyner}
\end{equation}
where $X\leftrightarrow W\leftrightarrow Y$ denotes a Markov chain,
i.e., $X$ and $Y$ are conditionally independent given $W$.

Denote the \emph{coupling set} of $(P_{X},P_{Y})$ as 
\begin{align}
\mathcal{C}(P_{X},P_{Y}) & :=\bigl\{ Q_{XY}\in\mathcal{P}(\mathcal{X}\times\mathcal{Y}):\nonumber \\
 & \qquad Q_{X}=P_{X},Q_{Y}=P_{Y}\bigr\}.
\end{align}
Define the \emph{maximal $s$-mixed Shannon-cross entropy} with respect
to $\pi_{XY}$ over couplings $\mathcal{C}(P_{X},P_{Y})$ as 
\begin{align}
 & H_{s}(P_{X},P_{Y}\|\pi_{XY})\nonumber \\
 & :=\max_{Q_{XY}\in\mathcal{C}(P_{X},P_{Y})}\sum_{x,y}Q_{XY}(x,y)\log\frac{1}{\pi\left(x,y\right)}+\frac{1}{s}H(Q_{XY}).\label{eq:maximalcrossentropy}
\end{align}
The \emph{$t$-relaxed Wyner's common information} of $\pi_{XY}$
\cite{sula2021common} is defined as 
\[
C(\pi_{XY},t):=\underset{Q_{W|XY}:I_{Q}(X;Y|W)\le t}{\inf}I_{Q}(XY;W).
\]
In particular, for $\mathrm{DSBS}(r)$, we denote its relaxed Wyner's
common information as $C(r,t)$.

\subsection{Problem Formulation}

\begin{figure}
\centering \setlength{\unitlength}{0.05cm} 
{ \begin{picture}(100,60) 
\put(5,30){\line(1,0){20}} \put(25,30){\vector(1,1){18}}
\put(25,30){\vector(1,-1){18}} \put(44,42){\framebox(30,12){$P_{X^{n}|M_{n}}$}}
\put(44,6){\framebox(30,12){$P_{Y^{n}|M_{n}}$}} \put(74,48){\vector(1,0){22}}
\put(74,12){\vector(1,0){22}} \put(10,33){%
\mbox{%
$M_{n}$%
}} \put(80,50){%
\mbox{%
$X^{n}$%
}} \put(80,14){%
\mbox{%
$Y^{n}$%
}} \end{picture}} \caption{Distributed source synthesis problem, where $M_{n}$ is a random variable
uniformly distributed over $\calM_{n}:=\{1,\ldots,2^{nR}\}$.}
\label{fig:dss} 
\end{figure}
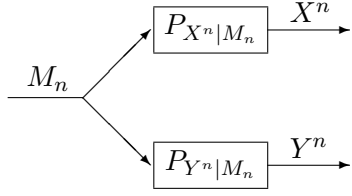

In this paper, we consider the distributed source simulation problem
illustrated in Fig.~\ref{fig:dss}. Given a target distribution $\pi_{XY}$,
we wish to minimize the alphabet size of a random variable $M_{n}$
that is uniformly distributed over\footnote{For simplicity, we assume that $2^{nR}$ and similar expressions are
integers.} $\calM_{n}:=\{1,\ldots,2^{nR}\}$ ($R$ is a positive number known
as the {\em rate}), such that the generated (or synthesized) distribution
\begin{align}
 & P_{X^{n}Y^{n}}(x^{n},y^{n})\nonumber \\
 & \qquad:=\frac{1}{|{\cal M}_{n}|}\sum_{m\in{\cal M}_{n}}P_{X^{n}|M_{n}}(x^{n}|m)P_{Y^{n}|M_{n}}(y^{n}|m)\label{eq:-113}
\end{align}
forms a good approximation to the product distribution $\pi_{XY}^{\otimes n}$.
The pair of random mappings $(P_{X^{n}|M_{n}},P_{Y^{n}|M_{n}})$ constitutes
a \emph{synthesis code}.

The minimum rates required to ensure these two measures vanish asymptotically
are respectively termed the\emph{ unnormalized and normalized Rényi
common information}, and denoted as for $\alpha\in[-\infty,\infty]$,
\begin{align}
 & T_{\alpha}(\pi_{XY})\nonumber \\
 & \quad:=\inf\left\{ R:\;\lim_{n\to\infty}D_{\alpha}(P_{X^{n}Y^{n}}\|\pi_{XY}^{\otimes n})=0\right\} ,\\
 & \widetilde{T}_{\alpha}(\pi_{XY})\nonumber \\
 & \quad:=\inf\Big\{ R:\;\lim_{n\to\infty}\frac{1}{n}D_{\alpha}(P_{X^{n}Y^{n}}\|\pi_{XY}^{\otimes n})=0\Big\}.
\end{align}
It is clear that 
\begin{equation}
\widetilde{T}_{\alpha}(\pi_{XY})\le T_{\alpha}(\pi_{XY}).\label{eqn:stronger}
\end{equation}

Rényi divergences admit the following properties \cite{Erven}. 
\begin{enumerate}
\item (Skew Symmetry). For $\alpha\in[-\infty,\infty]\backslash\{0,1\}$,
$D_{\alpha}(Q\|P)=\frac{\alpha}{1-\alpha}D_{1-\alpha}(P\|Q)$ for
probability measures $P,Q$. 
\item (Signs). For $\alpha\in[0,\infty]$, $D_{\alpha}(Q\|P)\ge0$ for probability
measures $P,Q$. For $\alpha\in[-\infty,0)$, $D_{\alpha}(Q\|P)\le0$
for probability measures $P,Q$. Moreover, $D_{\alpha}(Q\|P)=0$ for
some $\alpha\in[-\infty,\infty]\backslash\{0\}$ if and only if $P=Q$;
$D_{0}(Q\|P)=0$ if and only if $Q\ll P$. 
\item (Monotonicity). Given $P$ and $Q$, $D_{\alpha}(Q\|P)$ is nondecreasing
in $\alpha\in[-\infty,\infty]$. 
\end{enumerate}
By skew symmetry of the Rényi divergence, for $\alpha\in[-\infty,\infty]\backslash\{0,1\}$,
\begin{align}
 & T_{\alpha}(\pi_{XY})\nonumber \\
 & \quad=\inf\left\{ R:\;\lim_{n\to\infty}D_{1-\alpha}(\pi_{XY}^{\otimes n}\|P_{X^{n}Y^{n}})=0\right\} ,\\
 & \widetilde{T}_{\alpha}(\pi_{XY})\nonumber \\
 & \quad=\inf\Big\{ R:\;\lim_{n\to\infty}\frac{1}{n}D_{1-\alpha}(\pi_{XY}^{\otimes n}\|P_{X^{n}Y^{n}})=0\Big\}.
\end{align}
By the monotonicity and the signs, for $0\le\alpha<\beta$ or for
$\beta<\alpha\le0$, it holds that $T_{\alpha}(\pi_{XY})\le T_{\beta}(\pi_{XY})$
and $\widetilde{T}_{\alpha}(\pi_{XY})\le\widetilde{T}_{\beta}(\pi_{XY})$. 

\section{Main Results }

For $\alpha=1+s\in(1,\infty)$, define 
\begin{align}
\Gamma_{\alpha}^{\mathrm{UB}}(\pi_{XY}) & :=\min_{P_{W}P_{X|W}P_{Y|W}:P_{XY}=\pi_{XY}}-\frac{1+s}{s}H(XY|W)\nonumber \\
 & \qquad+\sum_{w}P(w)H_{s}(P_{X|W=w},P_{Y|W=w}\|\pi_{XY})\label{eq:-35-4}
\end{align}
and 
\begin{align}
\Gamma_{\alpha}^{\mathrm{LB}}(\pi_{XY}) & :=\inf_{P_{W}P_{X|W}P_{Y|W}:P_{XY}=\pi_{XY}}-\frac{1+s}{s}H(XY|W)\nonumber \\
 & \qquad+\inf_{Q_{WW'}\in\mathcal{C}(P_{W},P_{W})}\sum_{w,w'}Q(w,w')\nonumber \\
 & \qquad\times H_{s}(P_{X|W=w},P_{Y|W=w'}\|\pi_{XY}).\label{eq:-35-4-1}
\end{align}
Define $\Gamma_{1}^{\mathrm{UB}}(\pi_{XY}),\Gamma_{1}^{\mathrm{LB}}(\pi_{XY}),\Gamma_{\infty}^{\mathrm{UB}}(\pi_{XY}),$
and $\Gamma_{\infty}^{\mathrm{LB}}(\pi_{XY})$ as the continuous extensions
of $\Gamma_{\alpha}^{\mathrm{UB}}(\pi_{XY})$ and $\Gamma_{\alpha}^{\mathrm{LB}}(\pi_{XY})$
as $\alpha$ tends to $1$ or $\infty$.

For $\alpha\in[-\infty,0)$, define 
\begin{align}
\Gamma_{\alpha}^{\mathrm{UB}}(\pi_{XY}) & :=\underset{Q_{XY}}{\sup}\underset{\substack{Q_{W|XY}:\\
I_{Q}(X;Y|W)\le(1-1/\alpha)D(Q_{XY}\|\pi_{XY})
}
}{\inf}I_{Q}(XY;W)\nonumber \\
 & =\underset{Q_{XY}}{\sup}C(Q_{XY},(1-1/\alpha)D(Q_{XY}\|\pi_{XY})).\label{eq:-15}
\end{align}

We summarize results on Rényi common informations in the following
theorem. 
\begin{thm}[Rényi Common Informations]
\label{thm:RenyiCI} \cite{WynerCI,Hayashi06,YuTan2018,YuTan2020_exact,yu2020corrections,yu2024renyi}
For any distribution $\pi_{XY}$ on a finite alphabet, the following
hold. 
\begin{enumerate}
\item For $\alpha\in[0,1]$, 
\begin{align*}
\widetilde{T}_{\alpha}(\pi_{XY}) & =T_{\alpha}(\pi_{XY})=\begin{cases}
C_{\mathsf{W}}(\pi_{XY}), & (0,1]\\
0, & \alpha=0
\end{cases}.
\end{align*}
\item For $\alpha\in(1,\infty]$, 
\[
\Gamma_{\alpha}^{\mathrm{LB}}(\pi_{XY})\le\widetilde{T}_{\alpha}(\pi_{XY})\leq T_{\alpha}(\pi_{XY})\leq\Gamma_{\alpha}^{\mathrm{UB}}(\pi_{XY}).
\]
\item For $\alpha\in[-\infty,0)$, 
\begin{equation}
C_{\mathsf{W}}(\pi_{XY})\le\widetilde{T}_{\alpha}(\pi_{XY})\leq T_{\alpha}(\pi_{XY})\leq\Gamma_{\alpha}^{\mathrm{UB}}(\pi_{XY}).\label{eq:-9}
\end{equation}
\item Furthermore, for $\alpha\in(0,\infty]$, the optimal Rényi divergence
$D_{\alpha}(P_{X^{n}Y^{n}}\|\pi_{XY}^{\otimes n})$ in the definitions
of the Rényi common informations decays at least exponentially fast
in $n$ when the common randomness rate $R$ satisfies that $R>C_{\mathsf{W}}(\pi_{XY})$
for $\alpha\in(0,1]$ and $R>\Gamma_{\alpha}^{\mathrm{UB}}(\pi_{XY})$
for $\alpha\in[-\infty,0)\cup(1,\infty]$. 
\end{enumerate}
\end{thm}
The case of $\alpha=1$ was proven in \cite{WynerCI,Hayashi06} and
the case of $\alpha\in(0,1)\cup(1,2]\cup\{\infty\}$ was proven in
\cite{YuTan2018,yu2020corrections}. For the case of $\alpha\in(2,\infty)$,
the achievability part can be proven by using the typical code in
\cite{yu2020corrections} and applying \cite[Lemma 1]{yu2024renyi}
to analyze the performance of this code, and the converse was proven
in \cite{yu2020corrections}. The proof for the case of $\alpha\in[-\infty,0)$
is provided in Section \ref{sec:Proof-of-Theorem-1}.

We now consider a doubly symmetric binary source (DSBS), denoted by
$\mathrm{DSBS}(\epsilon)$, with joint distribution 
\begin{equation}
\pi_{XY}:=\left[\begin{array}{cc}
\frac{1-\epsilon}{2} & \frac{\epsilon}{2}\\
\frac{\epsilon}{2} & \frac{1-\epsilon}{2}
\end{array}\right],
\end{equation}
where $\epsilon\in[0,1/2]$ is called the crossover probability. That
is equivalent to the setting that $X\sim\mathrm{Bern}(\frac{1}{2})$
and $Y=X\oplus Z$ with $Z\sim\mathrm{Bern}(\epsilon)$ independent
of $X$, and also equivalent to the setting that $X=W\oplus U$, $Y=W\oplus V$,
where $W\sim\mathrm{Bern}(\frac{1}{2})$, $U\sim\mathrm{Bern}(a)$,
and $V\sim\mathrm{Bern}(a)$ are independent and $a=\frac{1-\sqrt{1-2\epsilon}}{2}$,
i.e., $\epsilon=2a(1-a)$. The analytic expression for the Rényi common
information for the DSBS is given in the following theorem. 

Define 
\[
\omega(\epsilon,s):=\log^{3}\left(1-s\right)+\frac{ds^{2}}{1-s}\left(\left(\frac{1-\epsilon}{1-2\epsilon}\right)^{2}s-1\right),
\]
where 
\[
d=4\left(1-H(b)\right)^{2}\frac{\left(1-\epsilon\right)^{3}}{1-2\epsilon}\log\frac{\bar{\epsilon}}{\epsilon}
\]
with $b:=\frac{a^{2}}{a^{2}+\bar{a}^{2}}=\frac{1-\epsilon-\sqrt{1-2\epsilon}}{2(1-\epsilon)}$.

\textbf{Condition 1: }$\omega(\epsilon,s)\le0$ for all $s\in[0,\frac{1-2\epsilon}{\left(1-\epsilon\right)^{2}}]$.

\begin{figure*}
\centering \includegraphics[width=0.6\textwidth]{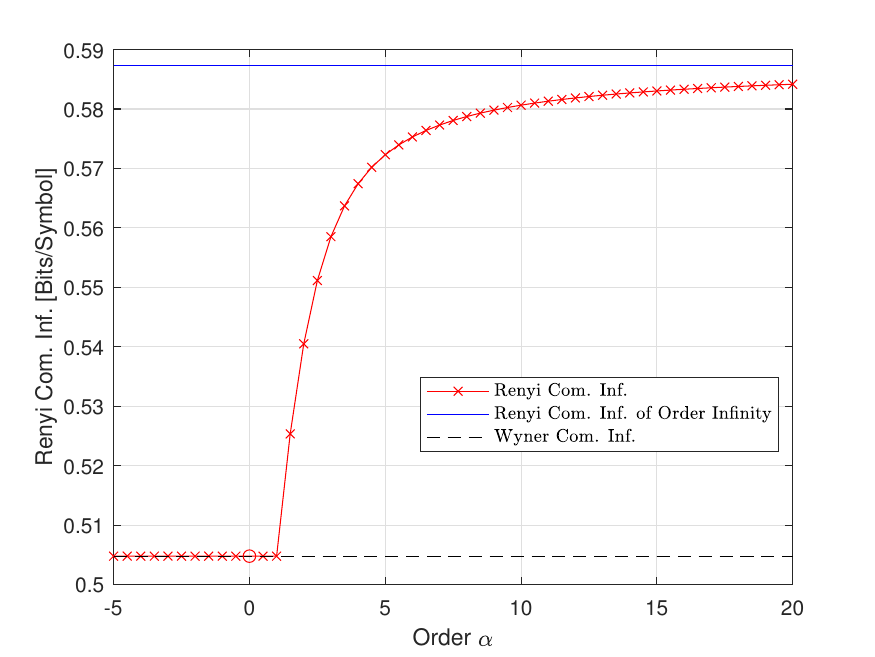}

\caption{\protect\label{fig:Common-informations-for}Illustrations of the Rényi
common informations for the DSBS with crossover probability $\epsilon=0.3$.
Wyner common information and the Rényi common information of order
$\infty$ respectively form a lower and an upper bound on the Rényi
common informations of all orders except order $0$. For order $0$,
the Rényi common information is zero.}
\end{figure*}

\begin{thm}[Rényi Common Information for DSBS]
\label{thm:For-a-DSBS} For the DSBS $\pi_{XY}=\mathrm{DSBS}(\epsilon)$
with $\epsilon\in[0,1/2]$, the following hold. 
\begin{enumerate}
\item For $\alpha\in[0,\infty]$, 
\begin{align}
\widetilde{T}_{\alpha}(\pi_{XY}) & =T_{\alpha}(\pi_{XY})=\Gamma_{\alpha}(\pi_{XY}),\label{eq:}
\end{align}
where 
\begin{align*}
\Gamma_{\alpha}(\pi_{XY}) & =\begin{cases}
0, & \alpha=0\\
1+H(\epsilon)-2H(a), & \alpha\in(0,1]\\
1-(1+2p^{*}-2a)\log(1-\epsilon)-(2a-2p^{*})\log\epsilon\\
\qquad-\frac{1+s}{s}2H(a)+\frac{1}{s}H(p^{*},a-p^{*},a-p^{*},1+p^{*}-2a), & \alpha\in(1,\infty)\\
1-(1-2a)\log(1-\epsilon)-2a\log\epsilon-2H(a), & \alpha=\infty
\end{cases}
\end{align*}
with $s=\alpha-1$, $p^{*}=\frac{-1+\sqrt{\kappa^{2}(1-2a)^{2}+4\kappa a(1-a)}}{2(\kappa-1)}-(\frac{1}{2}-a)$,
$\kappa=\left(\frac{1-\epsilon}{\epsilon}\right)^{2s}$, and $H(a_{1},a_{2},a_{3},a_{4})=-\sum_{i=1}^{4}a_{i}\log a_{i}$. 
\item Suppose that $\epsilon$ satisfies Condition 1. Then, for all $\alpha\in[-\infty,0)$,
\begin{align}
\widetilde{T}_{\alpha}(\pi_{XY}) & =T_{\alpha}(\pi_{XY})=1+H(\epsilon)-2H(a).\label{eq:-8}
\end{align}
\end{enumerate}
\end{thm}
Condition 1 is numerically verified to be true for $\epsilon\ge\epsilon_{0}$,
where $\epsilon_{0}\approx0.05510465170298144$. The Rényi common
information of various orders for the DSBS with crossover probability
$\epsilon=0.3$ is illustrated in Fig.~\ref{fig:Common-informations-for}.
From this figure, we can clearly see how the Rényi common information
transitions from the Wyner common information to the exact common
information. 

The case $\alpha\in[0,1]\cup\{\infty\}$ in the above theorem is not
new, which was proven in \cite{WynerCI,YuTan2018,yu2020corrections}.
Equation \eqref{eq:} for $\alpha\in(1,\infty)$ was previously conjectured
to be true by the present author and Tan in \cite{yu2020corrections}.
Theorem \ref{thm:For-a-DSBS} resolves this conjecture. Based on Theorem
\ref{thm:RenyiCI}, it is easy to upper bound $T_{\alpha}(\pi_{XY})$
and $\widetilde{T}_{\alpha}(\pi_{XY})$ by $\Gamma_{\alpha}(\pi_{XY})$,
as evaluated in \cite{yu2020corrections}. However, the other direction,
i.e., lower bounding $T_{\alpha}(\pi_{XY})$ and $\widetilde{T}_{\alpha}(\pi_{XY})$
by $\Gamma_{\alpha}(\pi_{XY})$, is not straightforward. The proof
of this case is provided in Section \ref{sec:Proof-of-Theorem}.

Equation \eqref{eq:-8} was proven by evaluating the upper and lower
bounds in \eqref{eq:-9} for the DSBS. The lower bound $C_{\mathsf{W}}(\pi_{XY})$,
i.e., Wyner common information, was already known to equal $1+H(\epsilon)-2H(a)$
for all $\epsilon\in[0,1/2]$ \cite{WynerCI}. A proof for the upper
bound is provided in Section \ref{sec:Proof-of-Theorem}. In this
proof, we use the analytic expression for the relaxed Wyner common
information to prove that $\Gamma_{-\infty}^{\mathrm{UB}}(\pi_{XY})=1+H(\epsilon)-2H(a)$
for $\epsilon$ satisfying Condition 1. By the monotonicity of the
Rényi divergence in its order, $\Gamma_{\alpha}^{\mathrm{UB}}(\pi_{XY})=1+H(\epsilon)-2H(a)$
for all $\alpha\in[-\infty,0)$ and for $\epsilon$ satisfying Condition
1. So, the upper and lower bounds coincide for this case. However,
numerical results show that Condition 1 is true if and only if $\epsilon\ge\epsilon_{0}$.
Numerical results also show that  $\Gamma_{-\infty}^{\mathrm{UB}}(\pi_{XY})>1+H(\epsilon)-2H(a)$
for all $\epsilon<\epsilon_{0}$. We believe that for an arbitrary
discrete source $\pi_{XY}$, the upper bound $\Gamma_{\alpha}^{\mathrm{UB}}(\pi_{XY})$
in \eqref{eq:-9} is tight. If so, there is a certain phase transition
occurring around $\epsilon=\epsilon_{0}$: For $\epsilon\ge\epsilon_{0}$,
the $-\infty$-Rényi common information is equal to the Wyner common
information; while for $\epsilon<\epsilon_{0}$, the $-\infty$-Rényi
common information is strictly larger than the Wyner common information.
We formally state this conjecture as follows. 

\begin{conjecture}
For sufficiently large $\beta$ and for $\pi_{XY}=\mathrm{DSBS}(\epsilon)$
with sufficiently small $\epsilon$, it holds that 
\begin{align}
\widetilde{T}_{-\beta}(\pi_{XY}) & >1+H(\epsilon)-2H(a).\label{eq:-8-1}
\end{align}
\end{conjecture}

The analytic expression for the relaxed Wyner common information is
given as follows. As mentioned above, this result will be used in
our proof of Theorem \ref{thm:For-a-DSBS}. 
\begin{prop}[Relaxed Wyner's Common Information for DSBS]
\label{prop:For-a-DSBS}For a DSBS $Q_{XY}=\mathrm{DSBS}(r)$ with
$r\in[0,1/2]$, the relaxed Wyner's common information of $Q_{XY}$
satisfies 
\begin{equation}
C(r,t)=\begin{cases}
\bar{r}\left(1-H(q)\right), & 0\le t<1-H(r)\\
0, & t\ge1-H(r)
\end{cases},\label{eq:-5}
\end{equation}
where $q$ is chosen as the solution to 
\begin{equation}
2H(\bar{r}q+\frac{r}{2})-\bar{r}H(q)-r-H(r)=t\label{eq:-7}
\end{equation}
such that $q\in[q_{0},1/2]$ with $q_{0}:=\frac{(1-\sqrt{1-2r})^{2}}{4\bar{r}}$.
An optimal $Q_{W|XY}$ attaining $C(r,s)$ is the one given in \eqref{eq:-4}
with $q$ chosen to satisfy \eqref{eq:-7}, which can be alternatively
expressed as the joint distribution of $W=W_{0}\oplus V_{0}$, $X=W_{0}\oplus V_{1}$,
and $Y=W_{0}\oplus V_{2}$ with $W_{0}\sim\mathrm{Bern}(\frac{1}{2})$,
$V_{0}\sim\mathrm{Bern}(b_{0})$, $V_{1}\sim\mathrm{Bern}(b)$ and
$V_{2}\sim\mathrm{Bern}(b)$ independent, where $b=\frac{1-\sqrt{1-2r}}{2}$
, $b_{0}=\frac{q-b}{1-2b}$, and $q$ is the solution to \eqref{eq:-7}
(i.e., $b*b=r$ and $q_{0}*b=q$ with $q$ chosen to satisfy $I_{Q}(X;Y|W)=t$).
Moreover, this distribution is the unique optimal conditional distribution
attaining $C(r,t)$ for $0<r<1/2$ and $0\le t<1-H(r)$. 
\end{prop}
The ``$\le$'' part of this result was proven in \cite{wang2015function,sula2021common},
and was conjectured therein to be an equality. Proposition \ref{prop:For-a-DSBS}
resolves their conjecture. In fact, the ``$\ge$'' part is a consequence
of the characterization of the mutual information region given in
our recent work \cite{yu2022gray}.

Specifically, the proof is as follows. Observe that 
\[
I_{Q}(X;Y|W)=I_{Q}(XY;W)-I_{Q}(X;W)-I_{Q}(Y;W)+I_{Q}(X;Y).
\]
Given $Q_{XY}$, $I_{Q}(X;Y)$ is fixed. So, if we denote $\mathcal{I}$
as the mutual information region defined by 
\[
\{(I_{Q}(X;W),I_{Q}(Y;W),I_{Q}(XY;W))\}_{Q_{W|XY}},
\]
then 
\[
C(r,t)=\underset{(\alpha,\beta,\gamma)\in\mathcal{I}:\gamma-\alpha-\beta\le t-I_{Q}(X;Y)}{\inf}\gamma,
\]
i.e., the relaxed Wyner's common information is determined by the
mutual information region. The explicit formula for the mutual information
region of the DSBS was given in our recent work \cite[Theorem 2]{yu2022gray}.
Using this formula, we can easily obtain Proposition \ref{prop:For-a-DSBS}.

\section{Discussion}

Common information admits an intimate connection with the nonnegative
rank, which provides a common information-theoretic perspective for
the nonnegative rank; see a review on this topic in \cite{yu2022common}.
 The {\em nonnegative rank} of a nonnegative matrix $\bM\in\bbR_{+}^{m\times k}$,
denoted as $\rank_{+}(\bM)$, is the smallest integer $r$ such that
$\bM$ can be represented as 
\begin{equation}
\bM=\bU\bV^{\top}\label{eqn:rank_one}
\end{equation}
for some nonnegative matrices $\bU\in\bbR_{+}^{m\times r}$ and $\bV\in\bbR_{+}^{k\times r}$.
In particular, regarding a joint distribution $\pi_{XY}^{\otimes n}$
as a matrix, the nonnegative rank of $\pi_{XY}^{\otimes n}$ is the
minimum of the cardinality of $\mathcal{W}$ such that 
\begin{equation}
\pi_{XY}^{\otimes n}=\sum_{w\in\mathcal{W}}P_{W}(w)P_{X^{n}|W=w}P_{Y^{n}|W=w}\label{eqn:rank_one-1}
\end{equation}
 for some distribution tuple $(P_{W},P_{X^{n}|W},P_{Y^{n}|W})$. That
is, $\rank_{+}(\pi_{XY}^{\otimes n})=2^{G_{0}(\pi_{XY}^{\otimes n})}$,
where 
\begin{equation}
G_{\alpha}(\pi_{XY}^{\otimes n}):=\min_{P_{W}P_{X^{n}|W}P_{Y^{n}|W}:P_{X^{n}Y^{n}}=\pi_{XY}^{\otimes n}}H_{\alpha}(W)\label{eqn:cmn_ent_alpha-1}
\end{equation}
is the {\em exact Rényi common information of order $\alpha$}.
Here $H_{\alpha}(W)$ denotes the Rényi entropy of $W$.  It is known
that $\lim_{n\to\infty}\frac{G_{1}(\pi_{XY}^{n})}{n}$ is equal to
the exact common information and also equal to the Rényi common information
of order $\infty$. By definition, $\rank_{+}(\pi_{XY}^{\otimes n})\ge\rank(\pi_{XY}^{\otimes n})=\rank(\pi_{XY})^{n}$
and $\rank_{+}(\pi_{XY}^{\otimes n})\ge G_{1}(\pi_{XY}^{n})$. 

The Rényi common information provides an approximate version of the
nonnegative rank. The Rényi common information $T_{\alpha}(\pi_{XY})$
(or $\widetilde{T}_{\alpha}(\pi_{XY})$) of $\pi_{XY}$ is the minimum
of $\frac{1}{n}\log|\mathcal{W}|$ such that $\sum_{w\in\mathcal{W}}P_{W}(w)P_{X^{n}|W}P_{Y^{n}|W}$
is close to $\pi_{XY}^{\otimes n}$ under the Rényi divergence for
some tuple $(P_{W},P_{X^{n}|W},P_{Y^{n}|W})$.   So, $2^{nT_{\alpha}(\pi_{XY})}$
can be regarded as a variant of the nonnegative rank, called the $\alpha$-approximate
nonnegative rank, for which the exact equality requirement replaced
by an approximate equality requirement. 

For the DSBS $\pi_{XY}=\mathrm{DSBS}(\epsilon)$ with $\epsilon\in[0,1/2)$,
$\pi_{XY}^{\otimes n}(x^{n},y^{n})=\left(\frac{1-\epsilon}{2}\right)^{n}\left(\frac{\epsilon}{1-\epsilon}\right)^{d(x^{n},y^{n})}$,
where $d(x^{n},y^{n})$ is the Hamming distance between $x^{n}$ and
$y^{n}$. It can be seen that $\rank_{+}(\pi_{XY}^{\otimes n})=\rank(\pi_{XY}^{\otimes n})=2^{n}$.
Moreover, from Theorem \ref{thm:For-a-DSBS}, the $\alpha$-approximate
nonnegative rank satisfies that 
\[
2^{nT_{\alpha}(\pi_{XY})}=2^{n(\Gamma_{\alpha}(\pi_{XY})+o(1))},\quad\alpha\in[0,\infty],
\]
and under Condition 1, 
\[
2^{nT_{\alpha}(\pi_{XY})}=2^{n(1+H(\epsilon)-2H(a)+o(1))},\quad\alpha\in[-\infty,0).
\]

\section{\protect\label{sec:Proof-of-Theorem-1}Proof of Theorem \ref{thm:RenyiCI}
for $\alpha\in[-\infty,0)$}

\subsection{Converse }

By the monotonicity and the skew symmetry of Rényi divergence, $\widetilde{T}_{\alpha}(\pi_{XY})\ge\widetilde{T}_{\beta}(\pi_{XY})$
holds for any $\alpha\in[-\infty,0)$ and $\beta\in(0,1)$. Moreover,
$\widetilde{T}_{\beta}(\pi_{XY})=C_{\mathsf{W}}(\pi_{XY})$. The converse
for $\alpha\in[-\infty,0)$ follows.

\subsection{\protect\label{subsec:Achievability-for-Normalized}Achievability
for Normalized Divergences }

We now prove the achievability part for $\alpha\in[-\infty,0)$. We
first prove that $\widetilde{T}_{\alpha}(\pi_{XY})\leq\Gamma_{\alpha}^{\mathrm{UB}}(\pi_{XY}).$
We follow the proof idea of Statement 3 of Theorem 1 in \cite{yu2024renyi}.

In this part, we use following notations. Denote $T_{w^{n}}$ as the
type (i.e., the empirical measure) of $w^{n}$. Denote $\mathcal{P}_{n}(\mathcal{W})$
as the set of $n$-types. Denote $\mathcal{P}_{n}(\mathcal{X}|w^{n})$
as the set of conditional distributions $T_{X|W}$ such that $T_{w^{n}}T_{X|W}$
is a joint $n$-type, and $\mathcal{P}_{n}(\mathcal{Y}|w^{n})$ is
defined similarly.

Let $R>3\epsilon>0$. Let $\mathcal{C}_{T_{W}}:=\{W^{n}(m)\}_{m\in[2^{nR}/|\mathcal{P}_{n}(\mathcal{W})|]}$
be a set of random sequences such that $W^{n}(m),m\in[2^{nR}/|\mathcal{P}_{n}(\mathcal{W})|]$
are drawn independently for different $m$'s and according to the
same distribution $\Unif(\mathcal{T}_{T_{W}})$. The rate of $\mathcal{C}_{T_{W}}$
is hence $R'=R-o_{n}(1)$. By Strong Packing-Covering Lemma in \cite[Lemma 10]{yu2024renyi},
there is a realization $c_{T_{W}}$ of $\mathcal{C}_{T_{W}}$ satisfying
that 
\begin{equation}
2^{n(R'-I_{T}(W;XY)-\epsilon)}\le|\mathcal{T}_{T_{W|XY}}(x^{n},y^{n})\cap c_{T_{W}}|\leq2^{n(R'-I_{T}(W;XY)+\epsilon)},\quad\forall(x^{n},y^{n})\in\mathcal{T}_{T_{XY}},\label{eq:-73-1}
\end{equation}
for all $T_{XY|W}$ such that $I_{T}(W;XY)\le R'-3\epsilon$. Here
$T_{W|XY}T_{XY}=T_{W}T_{XY|W}$.

Let $c:=\bigcup_{T_{W}}c_{T_{W}}$. Let $f:[2^{nR}]\to\mathcal{W}^{n}$
be the resolvability code based on $c$, i.e., the deterministic map
given by $f(m)=w^{n}(m)$ with $w^{n}(m)$ being the $m$-th codeword
in $c$.

Let $P_{X^{n}|W^{n}}$ and $P_{Y^{n}|W^{n}}$ be channels given by
\begin{align*}
P_{X^{n}|W^{n}=w^{n}} & =\frac{1}{|\mathcal{P}_{n}(\mathcal{X}|w^{n})|}\sum_{T_{X|W}\in\mathcal{P}_{n}(\mathcal{X}|w^{n})}\Unif(\mathcal{T}_{T_{X|W}}(w^{n})),\\
P_{Y^{n}|W^{n}=w^{n}} & =\frac{1}{|\mathcal{P}_{n}(\mathcal{Y}|w^{n})|}\sum_{T_{Y|W}\in\mathcal{P}_{n}(\mathcal{Y}|w^{n})}\Unif(\mathcal{T}_{T_{Y|W}}(w^{n})).
\end{align*}
Let $P_{X^{n}Y^{n}|W^{n}}=P_{X^{n}|W^{n}}P_{Y^{n}|W^{n}}$.

We denote $s=-\alpha>0$. Observe that 
\begin{align}
 & 2^{sD_{1+s}(\pi_{XY}^{\otimes n}\|P_{X^{n}Y^{n}})}\nonumber \\
 & =\sum_{x^{n},y^{n}}\pi^{1+s}(x^{n},y^{n})P^{-s}(x^{n},y^{n})\\
 & =\sum_{x^{n},y^{n}}\pi^{1+s}(x^{n},y^{n})(\sum_{m}2^{-nR}P(x^{n},y^{n}|w^{n}(m)))^{-s}\label{eq:-44}\\
 & =\sum_{T_{XY}}\sum_{(x^{n},y^{n})\in\mathcal{T}_{T_{XY}}}2^{(1+s)n\sum T_{XY}\log\pi_{XY}}(\sum_{T_{W|XY}}2^{-n(R+H_{T}(X|W)+H_{T}(Y|W)+o(1))}\cdot|\mathcal{T}_{T_{W|XY}}(x^{n},y^{n})\cap c|)^{-s}\\
 & \le\sum_{T_{XY}}\sum_{(x^{n},y^{n})\in\mathcal{T}_{T_{XY}}}2^{(1+s)n\sum T_{XY}\log\pi_{XY}}(\sum_{T_{W|XY}:I_{T}(W;XY)\le R'-3\epsilon}2^{-n(R+H_{T}(X|W)+H_{T}(Y|W)+o(1))}\cdot|\mathcal{T}_{T_{W|XY}}(x^{n},y^{n})\cap c|)^{-s}\\
 & \le\sum_{T_{XY}}\sum_{(x^{n},y^{n})\in\mathcal{T}_{T_{XY}}}2^{(1+s)n\sum T_{XY}\log\pi_{XY}}(\sum_{T_{W|XY}:I_{T}(W;XY)\le R'-3\epsilon}2^{n(R'-I_{T}(W;XY)-\epsilon-R-H_{T}(X|W)-H_{T}(Y|W)-o(1))})^{-s}\\
 & \doteq\max_{T_{XY}}\min_{T_{W|XY}:I_{T}(W;XY)\le R'-3\epsilon}2^{nH_{T}(XY)}2^{(1+s)n\sum T_{XY}\log\pi_{XY}}2^{sn(H_{T}(XY)+I_{T}(X;Y|W)+\epsilon+o(1))}\\
 & \doteq\max_{T_{XY}}\min_{T_{W|XY}:I_{T}(W;XY)\le R'-3\epsilon}2^{-(1+s)nD(T_{XY}\|\pi_{XY})+sn(I_{T}(X;Y|W)+\epsilon+o(1))}.\label{eq:-45}
\end{align}

Therefore, 
\begin{align}
\frac{1}{n}D_{1+s}(\pi_{XY}^{\otimes n}\|P_{X^{n}Y^{n}}) & \le\max_{T_{XY}}\min_{T_{W|XY}:I_{T}(W;XY)\le R-4\epsilon}I_{T}(X;Y|W)-\frac{1+s}{s}D(T_{XY}\|\pi_{XY})+\epsilon+o(1).\label{eq:-46}
\end{align}
By \cite[Lemma 14]{Tan11_IT} and letting $\epsilon\downarrow0$ yields
the desired upper bound for $q\in(1,\infty)$. 
\begin{align}
\frac{1}{n}D_{1+s}(\pi_{XY}^{\otimes n}\|P_{X^{n}Y^{n}}) & \le\max_{Q_{XY}}\min_{Q_{W|XY}:I_{Q}(W;XY)\le R}I_{Q}(X;Y|W)-\frac{1+s}{s}D(Q_{XY}\|\pi_{XY})+o(1).\label{eq:-46-1}
\end{align}
Note that $s=-\alpha$. It can be seen that if $R>\Gamma_{\alpha}^{\mathrm{UB}}(\pi_{XY})$,
then the upon bound turns out to be zero. That is, $\widetilde{T}_{\alpha}(\pi_{XY})\leq\Gamma_{\alpha}^{\mathrm{UB}}(\pi_{XY}).$

The desired upper bound for $\alpha=-\infty$ follows similarly.

\subsection{Exponential Behavior}

We next prove the exponential convergence for $R>\Gamma_{\alpha}^{\mathrm{UB}}(\pi_{XY})$
for $\alpha\in[-\infty,0)$, which implies $T_{\alpha}(\pi_{XY})\leq\Gamma_{\alpha}^{\mathrm{UB}}(\pi_{XY}).$
By assumption, $R>\Gamma_{\alpha}^{\mathrm{UB}}(\pi_{XY})$. Observe
that for any $R\ge\Gamma_{\alpha}^{\mathrm{UB}}(\pi_{XY})$, it holds
that 
\begin{equation}
\eta(R):=\max_{Q_{XY}}\eta(R,Q_{XY})\le0,\label{eq:-43}
\end{equation}
where 
\[
\eta(Q_{XY},R):=\min_{Q_{W|XY}:I_{Q}(W;XY)\le R}I_{Q}(X;Y|W)-(1-1/\alpha)D(Q_{XY}\|P_{XY}).
\]
By setting $Q_{XY}=P_{XY}$, 
\begin{align}
\eta(R) & \ge\min_{Q_{WXY}:Q_{XY}=\pi_{XY},\,I_{Q}(W;XY)\le R}I_{Q}(X;Y|W)\label{eq:-16}\\
 & \ge0.\nonumber 
\end{align}
So, for all $R\ge\Gamma_{\alpha}^{\mathrm{UB}}(\pi_{XY})$, it holds
that $\eta(R)=0$, and the maximum in \eqref{eq:-43} is attained
by $Q_{XY}=\pi_{XY}$.

We now make the following claim. Denote $\mathcal{B}_{r}(P):=\{Q:\|Q-P\|\le r\}$
as the ball at $P$ of radius $r$ under the total variation distance. 
\begin{claim}
\label{claim:For-,-the} For $R>\Gamma_{\alpha}^{\mathrm{UB}}(\pi_{XY})$,
it holds that $\eta(Q_{XY},R)\le-\epsilon_{\delta}$ for all $Q_{XY}\notin\mathcal{B}_{\delta/2}(\pi_{XY})$,
where $\epsilon_{\delta}>0$ is a term vanishing as\footnote{Here $\delta'$ does not denote the Hölder's conjugate of $\delta$.}
$\delta\downarrow0$. In particular, for $R>\Gamma_{\alpha}^{\mathrm{UB}}(\pi_{XY})$,
the optimization at the LHS of \eqref{eq:-43} is \emph{uniquely}
attained by $Q_{XY}=\pi_{XY}$. 
\end{claim}
\begin{IEEEproof}[Proof of Claim \ref{claim:For-,-the}]
We now prove the above claim. We assume $W$ defined on $\mathcal{X}\times\mathcal{Y}$.
By setting $Q_{W|XY}$ to the identify channel $\bone_{W|XY}$ (i.e.,
$W=(X,Y)$), we observe that for each $Q_{XY}$, the distribution
$Q_{W'XY}=Q_{XY}\bone_{W|XY}$ satisfies 
\begin{align*}
I_{Q}(X;Y|W) & =0.
\end{align*}

The condition in \eqref{eq:-43} implies that for all $Q_{XY}$, $\eta(Q_{XY},\Gamma_{\alpha}^{\mathrm{UB}}(\pi_{XY}))\le0,$
i.e., for all $Q_{XY}$, there is $Q_{W|XY}^{*}$ such that $I_{Q^{*}}(W;XY)\le\Gamma_{\alpha}^{\mathrm{UB}}(\pi_{XY})$
and $I_{Q^{*}}(X;Y|W)\le(1-1/\alpha)D(Q_{XY}\|\pi_{XY})$. Let $U\sim Q_{U}:=\Bern(\lambda)$
with $\lambda\in(0,1)$. Let $W'=(W,U)$. Let 
\[
Q_{W'XY}=Q_{U}Q_{XY}Q_{W|XYU},
\]
where 
\[
Q_{W|XYU}=\begin{cases}
Q_{W|XY}^{*}, & U=0\\
\bone_{W|XY}, & U=1
\end{cases}.
\]

For this new distribution $Q_{W'XY}$, 
\begin{align*}
I_{Q}(W';XY) & =(1-\lambda)I_{Q^{*}}(W;XY)+\lambda\log(|\mathcal{X}||\mathcal{Y}|)\\
 & \le(1-\lambda)\Gamma_{\alpha}^{\mathrm{UB}}(\pi_{XY})+\lambda\log(|\mathcal{X}||\mathcal{Y}|).
\end{align*}
That is, for $R>\Gamma_{\alpha}^{\mathrm{UB}}(\pi_{XY})$, there is
some $\lambda\in(0,1)$ such that $Q_{W'XY}$ satisfies $I_{Q}(W';XY)<R$.
Moreover, 
\begin{align*}
I_{Q}(X;Y|W')-(1-1/\alpha)D(Q_{XY}\|\pi_{XY}) & =(1-\lambda)I_{Q^{*}}(X;Y|W)-(1-1/\alpha)D(Q_{XY}\|\pi_{XY})\\
 & \le(1-\lambda)(1-1/\alpha)D(Q_{XY}\|\pi_{XY})-(1-1/\alpha)D(Q_{XY}\|\pi_{XY})\\
 & =-\lambda(1-1/\alpha)D(Q_{XY}\|\pi_{XY})\\
 & \le-\lambda(1-1/\alpha)\delta^{2}/2
\end{align*}
for all $Q_{XY}\notin\mathcal{B}_{\delta/2}(\pi_{XY})$, where the
last line follows by Pinsker's inequality.
\end{IEEEproof}
We now start to prove exponential convergence in Theorem \ref{thm:RenyiCI}.
Note that $R>I(W;XY)$ for some $P_{WXY}$ such that $P_{XY}=\pi_{XY}$
and $X\leftrightarrow W\leftrightarrow Y$, since $\Gamma_{\alpha}^{\mathrm{UB}}(\pi_{XY})\ge C_{\mathsf{W}}(\pi_{XY})$
for $q>1$. Let $\delta>0$ such that $R>\Gamma_{\alpha}^{\mathrm{UB}}(\pi_{XY})+\delta$.

We consider a typical code. Let $\tilde{\mathcal{C}}:=\{W^{n}(m)\}_{m\in[2^{nR}(1-2^{-n\delta})]}$
be a set of random sequences such that $W^{n}(m),m\in[2^{nR}(1-2^{-n\delta})]$
are drawn independently for different $m$'s and according to the
same distribution $Q_{W^{n}}:=P_{W}^{\otimes n}(\cdot|\mathcal{T}_{\delta}^{(n)}(P_{W}))$
with $\delta>0$. Consider a simulation code $(\tilde{\mathcal{C}},P_{X|W}^{\otimes n},P_{Y|W}^{\otimes n})$.
From \cite[Lemma 10]{yu2024renyi}, there is a realization $\tilde{c}$
of $\tilde{\mathcal{C}}$ satisfying that for some $\delta>\delta'>0$,
\[
\Big|\frac{\sum_{m\in[2^{nR}(1-2^{-n\delta})]}\theta_{m}(x^{n},y^{n})}{2^{nR}(1-2^{-n\delta})\mu}-1\Big|\leq2^{-n\epsilon},\;\forall(x^{n},y^{n})\in\mathcal{T}_{\delta'}^{(n)}(P_{XY}),
\]
where 
\[
\theta_{m}(x^{n},y^{n}):=P_{X|W}^{\otimes n}(x^{n}|W^{n}(m))P_{Y|W}^{\otimes n}(y^{n}|W^{n}(m))\bone\{(W^{n}(m),x^{n},y^{n})\in\mathcal{T}_{\delta}^{(n)}(P_{WXY})\}
\]
and 
\begin{align*}
\mu & :=\mathbb{E}_{\tilde{\mathcal{C}}}[\theta_{m}(x^{n},y^{n})]\in P_{XY}^{\otimes n}(x^{n},y^{n})(1\pm2^{-n\epsilon'})
\end{align*}
for some $\epsilon'>0$. So, 
\begin{align}
 & \frac{\sum_{m\in[2^{nR}(1-2^{-n\delta})]}\theta_{m}(x^{n},y^{n})}{2^{nR}(1-2^{-n\delta})}\ge(1-2^{-n\epsilon''})P_{XY}^{\otimes n}(x^{n},y^{n}),\;\forall(x^{n},y^{n})\in\mathcal{T}_{\delta'}^{(n)}(P_{XY}),
\end{align}
for some $\epsilon''>0$.

We now consider the code constructed in Section \ref{subsec:Achievability-for-Normalized}
of size $2^{n(R-\delta)}$, which is denoted as $(\hat{\mathcal{C}},P_{X^{n}|W^{n}},P_{Y^{n}|W^{n}})$.
Then, \eqref{eq:-73-1} still holds for some realization $\hat{c}$
of $\hat{\mathcal{C}}$ with $R'=R-\delta-o_{n}(1)$.

The final code used here is $c=\tilde{c}\cup\hat{c}$. We randomly
and uniformly select a codeword from $c$. If the selected codeword
is from $\tilde{c}$, then we use $P_{X|W}^{\otimes n}$ and $P_{Y|W}^{\otimes n}$
to generate $(X^{n},Y^{n})$. On the other hand, if the selected codeword
is from $\hat{c}$, then we use $P_{X^{n}|W^{n}}$ and $P_{Y^{n}|W^{n}}$
to generate $(X^{n},Y^{n})$. The rate of this code is $R$. Observe
that for this code $c$ and for $s=-\alpha>0$, 
\begin{align*}
2^{sD_{1+s}(\pi_{XY}^{\otimes n}\|Q_{X^{n}Y^{n}})} & =\sum_{x^{n},y^{n}}P^{1+s}(x^{n},y^{n})Q^{-s}(x^{n},y^{n})\\
 & =\Sigma_{1,n}+\Sigma_{2,n},
\end{align*}
where 
\begin{align*}
\Sigma_{1,n} & =\sum_{(x^{n},y^{n})\in\mathcal{T}_{\delta'}^{(n)}}P^{1+s}(x^{n},y^{n})(\sum_{m}2^{-nR}P(x^{n},y^{n}|w^{n}(m)))^{-s},\\
\Sigma_{2,n} & =\sum_{(x^{n},y^{n})\notin\mathcal{T}_{\delta'}^{(n)}}P^{1+s}(x^{n},y^{n})(\sum_{m}2^{-nR}P(x^{n},y^{n}|w^{n}(m)))^{-s}.
\end{align*}
We estimate the two sums above. For the first sum, we observe that
\begin{align*}
\Sigma_{1,n} & \le\sum_{(x^{n},y^{n})\in\mathcal{T}_{\delta'}^{(n)}}P^{1+s}(x^{n},y^{n})(2^{-nR}\sum_{m\in[2^{nR}(1-2^{-n\delta})]}P(x^{n},y^{n}|w^{n}(m))\bone\{(w^{n}(m),x^{n},y^{n})\in\mathcal{T}_{\delta}^{(n)}\})^{-s}\\
 & \le\sum_{(x^{n},y^{n})\in\mathcal{T}_{\delta'}^{(n)}}P^{1+s}(x^{n},y^{n})((1-2^{-n\delta})P(x^{n},y^{n})(1-2^{-n\epsilon''}))^{-s}\\
 & =\sum_{(x^{n},y^{n})\in\mathcal{T}_{\delta'}^{(n)}}P(x^{n},y^{n})(1-2^{-n\epsilon''})^{-s}(1-2^{-n\delta})^{-s}\\
 & =P(\mathcal{T}_{\delta'}^{(n)})(1-2^{-n\epsilon''})^{-s}(1-2^{-n\delta})^{-s}\\
 & \to1\textrm{ exponentially fast.}
\end{align*}

For the second sum, following steps similar to proof steps in \eqref{eq:-44}-\eqref{eq:-46},
we observe that 
\begin{align*}
\lim_{n\to\infty}\frac{1}{ns}\log\Sigma_{2,n} & \le\lim_{n\to\infty}\max_{T_{XY}\notin\mathcal{B}_{\delta'/2}}\min_{T_{W|XY}:I_{T}(W;XY)\le R-5\epsilon}I_{T}(X;Y|W)-q'D(T_{XY}\|P_{XY})+\epsilon.
\end{align*}
By \cite[Lemma 14]{Tan11_IT} and letting $\epsilon\downarrow0$ and
then by Claim \ref{claim:For-,-the}, we obtain that for $R-\delta>\Gamma_{\alpha}^{\mathrm{UB}}(\pi_{XY})$
(with $\delta$ chosen sufficiently small),

\begin{align*}
\lim_{n\to\infty}\frac{1}{ns}\log\Sigma_{2,n} & \le\max_{Q_{XY}\notin\mathcal{B}_{\delta'/2}}\min_{Q_{W|XY}:I_{Q}(W;XY)\le R}I_{Q}(X;Y|W)-q'D(Q_{XY}\|P_{XY})\\
 & \le-\epsilon_{\delta'}.
\end{align*}
That is, $\Sigma_{2,n}\to0$ exponentially fast.

Therefore, given $\delta$ and $\delta'$, 
\begin{align*}
2^{sD_{1+s}(\pi_{XY}^{\otimes n}\|Q_{X^{n}Y^{n}})} & \le1+2^{-n(\epsilon'''+o_{n}(1))}
\end{align*}
for some $\epsilon'''>0$, i.e., for $\alpha<0$, $D_{\alpha}(Q_{X^{n}Y^{n}}\|\pi_{XY}^{\otimes n})\to0$
exponentially fast.

\section{\protect\label{sec:Proof-of-Theorem}Proof of Theorem \ref{thm:For-a-DSBS} }

\subsection{Upper Bound for $\alpha\in(1,\infty)$}

Denote $\alpha=1+s$. Set $X=W\oplus U$, $Y=W\oplus V$, where $W\sim\mathrm{Bern}(\frac{1}{2})$,
$U\sim\mathrm{Bern}(a)$, and $V\sim\mathrm{Bern}(a)$ are independent.
For this choice, 
\begin{align}
H_{s}(P_{X|W=w},P_{Y|W=w}\|\pi_{XY}) & =-p^{*}\log p^{*}-2(a-p^{*})\log(a-p^{*})\nonumber \\
 & \qquad-(1+p^{*}-2a)\log(1+p^{*}-2a)\nonumber \\
 & \qquad-s(1+2p^{*}-2a)\log\frac{1-\epsilon}{2}-s(2a-2p^{*})\log\frac{\epsilon}{2}.
\end{align}
Hence, 
\begin{align}
\Gamma_{1+s}^{\mathrm{UB}}(\pi_{XY}) & \leq-\frac{1+s}{s}2H(a)+\frac{1}{s}\big\{-p^{*}\log p^{*}-2(a-p^{*})\log(a-p^{*})\nonumber \\
 & \qquad-(1+p^{*}-2a)\log(1+p^{*}-2a)\nonumber \\
 & \qquad-s(1+2p^{*}-2a)\log\frac{1-\epsilon}{2}-s(2a-2p^{*})\log\frac{\epsilon}{2}\big\}.
\end{align}

\subsection{Lower Bound for $\alpha\in(1,\infty)$}

Denote 
\begin{align}
\gamma_{1}(w) & :=\mathbb{P}\{X=0|W=w\}\\
\gamma_{2}(w) & :=\mathbb{P}\{Y=0|W=w\}.
\end{align}
Hence, 
\begin{align}
\mathbb{E}\gamma_{1}(W) & =\mathbb{P}\{X=0\}=\frac{1}{2}\\
\mathbb{E}\gamma_{2}(W) & =\mathbb{P}\{Y=0\}=\frac{1}{2}\\
\mathbb{E}[\gamma_{1}(W)\gamma_{2}(W)] & =\mathbb{P}\{X=0,Y=0\}=\frac{1-\epsilon}{2}.
\end{align}
Every $Q_{XY}\in\mathcal{C}(P_{X|W=w},P_{Y|W=w'})$ can be written
as 
\[
Q_{XY}=\left[\begin{array}{cc}
p & \gamma_{1}(w)-p\\
\gamma_{2}(w')-p & 1+p-\gamma_{1}(w)-\gamma_{2}(w')
\end{array}\right].
\]

By definition, 
\begin{align}
H_{s}(P_{X|W=w},P_{Y|W=w'}\|\pi_{XY}) & =g(\gamma_{1}(w),\gamma_{2}(w')),
\end{align}
where 
\begin{align}
g(\gamma_{1},\gamma_{2}) & :=\max_{\max\{0,\gamma_{1}+\gamma_{2}-1\}\le p\le\min\{\gamma_{1},\gamma_{2}\}}f(\gamma_{1},\gamma_{2},p)\label{eq:-55}
\end{align}
with 
\begin{align}
f(\gamma_{1},\gamma_{2},p) & :=-p\log p-(\gamma_{1}-p)\log(\gamma_{1}-p)-(\gamma_{2}-p)\log(\gamma_{2}-p)\nonumber \\
 & \qquad-(1+p-\gamma_{1}-\gamma_{2})\log(1+p-\gamma_{1}-\gamma_{2})\nonumber \\
 & \qquad-s(1+2p-\gamma_{1}-\gamma_{2})\log\frac{1-\epsilon}{2}-s(\gamma_{1}+\gamma_{2}-2p)\log\frac{\epsilon}{2}.
\end{align}
Due to the concavity of the entropy, the optimization above is a convex
optimization problem. So, we can easily solve it and obtain that 
\begin{align*}
g(\gamma_{1},\gamma_{2}) & =f(\gamma_{1},\gamma_{2},p^{*}(\gamma_{1},\gamma_{2})),
\end{align*}
where 
\begin{align*}
p^{*}(\gamma_{1},\gamma_{2}) & =\frac{\sqrt{\kappa^{2}\left(1-\gamma_{1}-\gamma_{2}\right)^{2}+\left(\gamma_{1}-\gamma_{2}\right)^{2}+2\kappa\left(\gamma_{1}(1-\gamma_{1})+\gamma_{2}(1-\gamma_{2})\right)}-1}{2\left(\kappa-1\right)}-\frac{1-\gamma_{1}-\gamma_{2}}{2}.
\end{align*}
Denote 
\begin{align}
g(\gamma_{1}) & :=g(\gamma_{1},\gamma_{1}).\label{eq:-55-3}
\end{align}
It is easy to see that $g(\gamma_{1},\gamma_{2})=g(1-\gamma_{1},1-\gamma_{2})$
and $g(\gamma_{1})=g(1-\gamma_{1})$.

The following is our main technical lemma. 
\begin{lem}[Splitting of Entropy]
\label{lem:entropy} Let $\gamma_{1},\gamma_{2},p_{1},p_{2}\in[0,1]$
be such that $\max\{0,2\gamma_{1}-1\}\le p_{1}\le\gamma_{1}$ and
$\max\{0,2\gamma_{2}-1\}\le p_{2}\le\gamma_{2}$. Then, the following
hold. 
\begin{enumerate}
\item It holds that 
\begin{align}
 & H(\gamma_{1})+H(\gamma_{2})\nonumber \\
 & \ge\frac{1}{2}\big(H(p_{1},\gamma_{1}-p_{1},\gamma_{1}-p_{1},1+p_{1}-2\gamma_{1})\nonumber \\
 & \qquad+H(p_{2},\gamma_{2}-p_{2},\gamma_{2}-p_{2},1+p_{2}-2\gamma_{2})\big).\label{eq:-59}
\end{align}
\item If $p:=\frac{p_{1}+p_{2}}{2}\le\gamma_{1}\gamma_{2}$, then it holds
that 
\begin{align}
 & H(p,\gamma_{1}-p,\gamma_{2}-p,1+p-\gamma_{1}-\gamma_{2})\nonumber \\
 & \ge\frac{1}{2}\big(H(p_{1},\gamma_{1}-p_{1},\gamma_{1}-p_{1},1+p_{1}-2\gamma_{1})\nonumber \\
 & \qquad+H(p_{2},\gamma_{2}-p_{2},\gamma_{2}-p_{2},1+p_{2}-2\gamma_{2})\big).\label{eq:-58}
\end{align}
\end{enumerate}
\end{lem}
By Lemma \ref{lem:entropy}, 
\begin{align*}
 & \frac{g(\gamma_{1})+g(\gamma_{2})}{2}+s\log\frac{1-\epsilon}{2}\\
 & =\max_{\substack{\max\{0,2\gamma_{1}-1\}\le p_{1}\le\gamma_{1}\\
\max\{0,2\gamma_{2}-1\}\le p_{2}\le\gamma_{2}
}
}\frac{1}{2}\big(H(p_{1},\gamma_{1}-p_{1},\gamma_{1}-p_{1},1+p_{1}-2\gamma_{1})+(\gamma_{1}+\gamma_{2}-2p_{1})\log\kappa\\
 & \qquad+H(p_{2},\gamma_{2}-p_{2},\gamma_{2}-p_{2},1+p_{2}-2\gamma_{2})+(\gamma_{1}+\gamma_{2}-2p_{2})\log\kappa\big)\\
 & \le\begin{cases}
H(p,\gamma_{1}-p,\gamma_{2}-p,1+p-\gamma_{1}-\gamma_{2})+(\gamma_{1}+\gamma_{2}-2p)\log\kappa, & \frac{p_{1}^{*}+p_{2}^{*}}{2}\le\gamma_{1}\gamma_{2}\\
H(\gamma_{1})+H(\gamma_{2})+(\gamma_{1}+\gamma_{2}-2\gamma_{1}\gamma_{2})\log\kappa, & \frac{p_{1}^{*}+p_{2}^{*}}{2}>\gamma_{1}\gamma_{2}
\end{cases}\\
 & \le\max_{\max\{0,\gamma_{1}+\gamma_{2}-1\}\le p\le\min\{\gamma_{1},\gamma_{2}\}}H(p,\gamma_{1}-p,\gamma_{2}-p,1+p-\gamma_{1}-\gamma_{2})+(\gamma_{1}+\gamma_{2}-2p)\log\kappa\\
 & =g(\gamma_{1},\gamma_{2})+s\log\frac{1-\epsilon}{2}.
\end{align*}
Hence, 
\begin{align*}
H_{s}(P_{X|W=w},P_{Y|W=w'}\|\pi_{XY}) & =\mathbb{E}_{W}\left[g(\gamma_{1}(W),\gamma_{2}(W'))\right]\\
 & \ge\frac{1}{2}\mathbb{E}_{W}\left[g(\gamma_{1}(W))+g(\gamma_{2}(W))\right],
\end{align*}
which implies 
\begin{align}
\Gamma_{1+s}^{\mathrm{LB}}(\pi_{XY}) & \geq\inf_{\substack{P_{W},\gamma_{1}(\cdot),\gamma_{2}(\cdot):\\
\mathbb{E}\gamma_{1}(W)=\frac{1}{2}\\
\mathbb{E}\gamma_{2}(W)=\frac{1}{2}\\
\mathbb{E}[\gamma_{1}(W)\gamma_{2}(W)]=\frac{1-\epsilon}{2}
}
}-\frac{1+s}{s}\mathbb{E}\left[H(\gamma_{1}(W))+H(\gamma_{2}(W))\right]\nonumber \\
 & \qquad+\frac{1}{2s}\mathbb{E}\left[g(\gamma_{1}(W))+g(\gamma_{2}(W))\right].
\end{align}

Define $\hat{\gamma}_{1}(W):=\left|\gamma_{1}(W)-\frac{1}{2}\right|$
and $\hat{\gamma}_{2}(W):=\left|\gamma_{2}(W)-\frac{1}{2}\right|$.
Define 
\begin{align*}
\chi_{s}(t) & :=-\frac{1+s}{s}2H\left(\frac{1}{2}-\sqrt{t}\right)+\frac{1}{s}g\left(\frac{1}{2}-\sqrt{t}\right)\\
 & =-\frac{1+s}{s}2H\left(\frac{1}{2}+\sqrt{t}\right)+\frac{1}{s}\Big(-(c+\sqrt{t})\log(c+\sqrt{t})-2(\frac{1}{2}-c)\log(\frac{1}{2}-c)\\
 & \qquad-(c-\sqrt{t})\log(c-\sqrt{t})-c\log\kappa-s\log\frac{1-\epsilon}{2}\Big),
\end{align*}
where 
\begin{align*}
c & :=\frac{\sqrt{\kappa+4\kappa(\kappa-1)t}-1}{2(\kappa-1)}.
\end{align*}
Then we can lower bound $\Gamma_{1+s}^{\mathrm{LB}}(\pi_{XY})$ as
\begin{align}
\Gamma_{1+s}^{\mathrm{LB}}(\pi_{XY}) & \geq\inf_{\substack{P_{W},\hat{\gamma}_{1}(\cdot),\hat{\gamma}_{2}(\cdot):\\
0\leq\hat{\gamma}_{1}(W),\hat{\gamma}_{2}(W)\leq\frac{1}{2}\\
\mathbb{E}[\hat{\gamma}_{1}(W)\hat{\gamma}_{2}(W)]\geq\frac{1-2\epsilon}{4}
}
}-\frac{1+s}{s}\mathbb{E}\left[H\left(\frac{1}{2}-\hat{\gamma}_{1}(W)\right)+H\left(\frac{1}{2}-\hat{\gamma}_{2}(W)\right)\right]\nonumber \\
 & \qquad+\frac{1}{2s}\mathbb{E}\left[g\left(\frac{1}{2}-\hat{\gamma}_{1}(W)\right)+g\left(\frac{1}{2}-\hat{\gamma}_{2}(W)\right)\right]\label{eq:-8-1-2}\\
 & =\inf_{\substack{P_{W},\hat{\gamma}_{1}(\cdot),\hat{\gamma}_{2}(\cdot):\\
0\leq\hat{\gamma}_{1}(W),\hat{\gamma}_{2}(W)\leq\frac{1}{2}\\
\mathbb{E}[\hat{\gamma}_{1}(W)\hat{\gamma}_{2}(W)]\geq\frac{1-2\epsilon}{4}
}
}\frac{1}{2}\mathbb{E}\left[\chi_{s}(\hat{\gamma}_{1}(W)^{2})+\chi_{s}(\hat{\gamma}_{2}(W)^{2})\right].
\end{align}

\begin{lem}
\label{lem:convexity}Given any $\kappa\ge1$ and $s>0$, $\chi_{s}(t)$
is convex and nondecreasing in $t\in[0,1/4]$. 
\end{lem}
By this lemma, 
\begin{align*}
\frac{1}{2}\mathbb{E}\left[\chi_{s}(\hat{\gamma}_{1}(W)^{2})+\chi_{s}(\hat{\gamma}_{2}(W)^{2})\right] & \ge\chi_{s}\left(\mathbb{E}\left[\frac{\hat{\gamma}_{1}(W)^{2}+\hat{\gamma}_{2}(W)^{2}}{2}\right]\right)\\
 & \ge\chi_{s}\left(\mathbb{E}[\hat{\gamma}_{1}(W)\hat{\gamma}_{2}(W)]\right).
\end{align*}
So, we can lower bound $\Gamma_{1+s}^{\mathrm{LB}}(\pi_{XY})$ as
\begin{align*}
\Gamma_{1+s}^{\mathrm{LB}}(\pi_{XY}) & \geq\chi_{s}\left(\frac{1-2\epsilon}{4}\right)=\Gamma_{1+s}(\pi_{XY}).
\end{align*}

\subsection{Upper Bound for $\alpha\in[-\infty,0)$}

The key in our proof is the following expression for $\Gamma_{\alpha}^{\mathrm{UB}}(\pi_{XY})$
for all $\epsilon\in[0,1/2]$. 
\begin{prop}
\label{prop:upperbound}For $\pi_{XY}=\mathrm{DSBS}(\epsilon)$ with
$\epsilon\in[0,1/2]$ and for $\alpha\in[-\infty,0)$, it holds that
\begin{align*}
\Gamma_{\alpha}^{\mathrm{UB}}(\pi_{XY}) & =\underset{r\in[0,\epsilon]}{\sup}C(r,(1-1/\alpha)D(r\|\epsilon)).
\end{align*}
\end{prop}
In other words, this proposition states that for the DSBS $\pi_{XY}$,
the distribution $Q_{XY}$ in the supremization in \eqref{eq:-15}
can be restricted to a DSBS.

To prove the desired result, it suffices to prove that $\Gamma_{-\infty}^{\mathrm{UB}}(\pi_{XY})=1+H(\epsilon)-2H(a)$
for $\epsilon\ge0.056$. By the proposition above, we only need to
show that for $\epsilon\ge0.056$, 
\[
\underset{r\in[0,\epsilon]}{\sup}C(r,D(r\|\epsilon))=1+H(\epsilon)-2H(a),
\]
i.e., the supremum is attained at $r=\epsilon$. By Proposition \ref{prop:For-a-DSBS},
it suffices to prove that for all $r\in[0,\epsilon]$, 
\[
\bar{r}\left(1-H(q)\right)\le1+H(\epsilon)-2H(a),
\]
where $q\in[q_{0},1/2]$ is the solution to \eqref{eq:-7}. This is
equivalent to showing that for all $q\in[b,1/2]$ with $b:=\frac{a^{2}}{a^{2}+\bar{a}^{2}}=\frac{1-\epsilon-\sqrt{1-2\epsilon}}{2(1-\epsilon)}$,
\begin{equation}
2H(\bar{r}q+\frac{r}{2})-\bar{r}H(q)-r-H(r)\le D(r\|\epsilon)\label{eq:-7-2-1}
\end{equation}
where $r$ is chosen as the solution to 
\begin{equation}
\bar{r}\left(1-H(q)\right)=\eta_{\epsilon},\label{eq:-5-1-3}
\end{equation}
with 
\[
\eta:=1+H(\epsilon)-2H(a)=\left(1-\epsilon\right)\left(1-H(b)\right).
\]

Solving $r$ from \eqref{eq:-5-1-3} and substituting it into \eqref{eq:-7-2-1},
we obtain the following equivalent inequality: for all $q\in[b,1/2]$,
\[
f(q):=2H\left(-\frac{\eta}{1-H(q)}\left(\frac{1}{2}-q\right)+\frac{1}{2}\right)-\frac{\eta}{1-H(q)}\log\frac{\epsilon}{\bar{\epsilon}}+\log\epsilon+\eta-1\le0,
\]
i.e., $f(q)\le f(b)$ since $f(b)=0$.

To prove $f(q)\le f(b)$, it suffices to $f'(q)\le0$ for $q\in[b,1/2]$.
That is, 
\begin{equation}
\left(\log\frac{1-H(q)+\eta\left(1-2q\right)}{1-H(q)-\eta\left(1-2q\right)}\right)\left(2+\log q+\log\left(1-q\right)\right)-\left(\log\left(1-q\right)-\log q\right)\log\frac{\epsilon}{\bar{\epsilon}}\le0.\label{eq:-3}
\end{equation}
Denote $q=\frac{1-t}{2}\ge b$. Then, \eqref{eq:-3} is equivalent
to that for all $t\in[0,c]$ with $c:=1-2b=\frac{\sqrt{1-2\epsilon}}{1-\epsilon}$,
\begin{equation}
g(t):=\log\frac{1-H(\frac{1-t}{2})+\eta t}{1-H(\frac{1-t}{2})-\eta t}+\frac{\log\left(1+t\right)-\log\left(1-t\right)}{\log\left(1-t^{2}\right)}\log\frac{\bar{\epsilon}}{\epsilon}\ge0.\label{eq:-1}
\end{equation}

The derivative of $g$ is as follows: 
\[
g'(t)=\frac{\eta\log\left(1-t^{2}\right)}{\left(1-H(\frac{1-t}{2})\right)^{2}-\eta^{2}t^{2}}+\frac{4\left(1-H(\frac{1-t}{2})\right)}{\left(1-t^{2}\right)\log^{2}\left(1-t^{2}\right)}\log\frac{\bar{\epsilon}}{\epsilon}.
\]
Note that $g(c)=0$. So, \eqref{eq:-1} is equivalent to that $g'(t)\le0$
for all $0\le t\le c$, i.e., 
\begin{equation}
\eta\left(1-t^{2}\right)\log^{3}\left(1-t^{2}\right)+4\left(1-H(\frac{1-t}{2})\right)\left(\left(1-H(\frac{1-t}{2})\right)^{2}-\eta^{2}t^{2}\right)\log\frac{\bar{\epsilon}}{\epsilon}\le0.\label{eq:-17}
\end{equation}

\begin{lem}
For any $c\in(0,1]$, it holds that 
\[
1-H(\frac{1-t}{2})\le\left(1-H(\frac{1-c}{2})\right)\left(\frac{t}{c}\right)^{2},\quad\forall t\in[0,c].
\]
That is, 
\[
\frac{1}{t^{2}}D(\frac{1-t}{2}\|\frac{1}{2})\le\frac{1}{c^{2}}D(\frac{1-c}{2}\|\frac{1}{2}),\quad\forall t\in[0,c],
\]
or equivalently, $\frac{1}{t^{2}}D(\frac{1-t}{2}\|\frac{1}{2})$ is
nondecreasing in $t\in(0,1]$. 
\end{lem}
\begin{IEEEproof}
Define $\varphi(t):=\frac{1}{t^{2}}D(\frac{1-t}{2}\|\frac{1}{2}).$
Then, $\varphi'(t)=\frac{\psi(t)}{t^{3}\ln2},$ where $\psi(t):=-(2-t)\log(1-t)-(t+2)\log(t+1)$.
It is easy to see that 
\[
\psi'(t)=\frac{1}{\ln2}\left(\frac{2t}{1-t^{2}}-\log\left(\frac{1+t}{1-t}\right)\right)\ge0.
\]
So, $\psi(t)\ge\psi(0)=0$ for $t\ge0$, which implies $\varphi'(t)\ge0$.
Hence, $\varphi(t)$ is nondecreasing in $t\in(0,1]$.
\end{IEEEproof}
By this lemma, for all $s=t^{2}\le\frac{1-2\epsilon}{\left(1-\epsilon\right)^{2}}$,
\eqref{eq:-17} is implied by the following inequality: 
\[
\omega(s):=\log^{3}\left(1-s\right)+\frac{ds^{2}}{1-s}\left(\left(\frac{1-\epsilon}{1-2\epsilon}\right)^{2}s-1\right)\le0,
\]
where 
\[
d=4\eta^{2}\frac{1-\epsilon}{1-2\epsilon}\log\frac{\bar{\epsilon}}{\epsilon}.
\]

\appendices{}

\section{Proof of Lemma \ref{lem:entropy} }

Statement 1 just follows from the fact that the independent coupling
of two random variables have largest joint entropy, i.e., $H(X,Y)\le H(X)+H(Y)$.
We next prove Statement 2.

Denote 
\[
p_{1}=p+s;\;p_{2}=p-s;\;\gamma_{1}=\gamma+t;\;\gamma_{2}=\gamma-t,
\]
and 
\begin{align}
\varphi(s,t) & :=H(p,\gamma_{1}-p,\gamma_{2}-p,1+p-\gamma_{1}-\gamma_{2})\nonumber \\
 & \qquad-\frac{1}{2}\big\{ H(p_{1},\gamma_{1}-p_{1},\gamma_{1}-p_{1},1+p_{1}-2\gamma_{1})\nonumber \\
 & \qquad+H(p_{2},\gamma_{2}-p_{2},\gamma_{2}-p_{2},1+p_{2}-2\gamma_{2})\big\}\label{eq:-13}\\
 & =-p\log(p)-(\gamma-p-t)\log(\gamma-p-t)\nonumber \\
 & \qquad-(\gamma-p+t)\log(\gamma-p+t)-(1+p-2\gamma)\log(1+p-2\gamma)\nonumber \\
 & \qquad+\frac{1}{2}\big\{(p-s)\log(p-s)+(p+s)\log(p+s)\nonumber \\
 & \qquad+2(\gamma-p+s-t)\log(\gamma-p+s-t)+2(\gamma-p-s+t)\log(\gamma-p-s+t)\nonumber \\
 & \qquad+(1+p-2\gamma-s+2t)\log(1+p-2\gamma-s+2t)\nonumber \\
 & \qquad+(1+p-2\gamma+s-2t)\log(1+p-2\gamma+s-2t)\big\}.\nonumber 
\end{align}
Given $(p,\gamma)\in[0,1]^{2}$ such that $\max\{0,2\gamma-1\}\le p\le\gamma^{2}$,
define 
\begin{align*}
\mathcal{R} & :=\big\{(s,t)\in[-p,p]\times[-\gamma,\gamma]:\\
 & \qquad\max\{0,2(\gamma+t)-1\}\le p+s\le\gamma+t,\\
 & \qquad\max\{0,2(\gamma-t)-1\}\le p-s\le\gamma-t,\\
 & \qquad t^{2}\le\gamma^{2}-p\big\}.
\end{align*}
To prove Statement 2, it suffices to show that $\varphi(s,t)\ge0$
for all $(s,t)\in\mathcal{R}$.

The set $\mathcal{R}$ is compact. So, the minimum of $\varphi(s,t)$
over $(s,t)\in\mathcal{R}$ is attained at some point $(s^{*},t^{*})$.
The point $(s^{*},t^{*})$ is either in the interior of $\mathcal{R}$
or at the boundary of $\mathcal{R}$.

We first assume $(s^{*},t^{*})$ is in the interior of $\mathcal{R}$.
For this case, $(s^{*},t^{*})$ must be a stationary point of $\varphi$.
In other words, $(s^{*},t^{*})$ satisfies the following equations:
\begin{align}
\partial_{s}\varphi(s,t) & =\frac{1}{2}\log\frac{(p+s)(\gamma-p+s-t)^{2}(1+p-2\gamma+s-2t)}{(p-s)(\gamma-p-s+t)^{2}(1+p-2\gamma-s+2t)}=0\\
\partial_{t}\varphi(s,t) & =\log\frac{(\gamma-p-t)(\gamma-p-s+t)(1+p-2\gamma-s+2t)}{(\gamma-p+t)(\gamma-p+s-t)(1+p-2\gamma+s-2t)}=0.
\end{align}
Solving these equations, we obtain the following solutions: For $\gamma\neq\frac{1}{2}$
and $\gamma\neq p+\frac{1}{4}$, 
\begin{align*}
s_{1} & =t_{1}=0,\\
s_{2} & =2\left(\gamma-\frac{\left|2\gamma-1\right|}{2\gamma-1}\sqrt{\gamma^{2}-p}\right)t_{2},\\
t_{2} & =\sqrt{\frac{(\gamma-p)\sqrt{\gamma^{2}-p}}{2\sqrt{\gamma^{2}-p}-\left|2\gamma-1\right|}},\\
s_{3} & =2\left(\gamma+\frac{\left|2\gamma-1\right|}{2\gamma-1}\sqrt{\gamma^{2}-p}\right)t_{3},\\
t_{3} & =\sqrt{\frac{(\gamma-p)\sqrt{\gamma^{2}-p}}{2\sqrt{\gamma^{2}-p}+\left|2\gamma-1\right|}},
\end{align*}
$(s_{4},t_{4})=-(s_{2},t_{2})$, and $(s_{5},t_{5})=-(s_{3},t_{3})$;
for $\gamma=\frac{1}{2}$, 
\begin{align*}
s_{1}' & =t_{1}'=0,\\
s_{2}' & =\frac{1}{2}\left(\sqrt{1-2p}-\sqrt{8p^{2}-6p+1}\right),\\
t_{2}' & =\frac{1}{2}\sqrt{1-2p},\\
s_{3}' & =\frac{1}{2}\left(\sqrt{1-2p}+\sqrt{8p^{2}-6p+1}\right),\\
t_{3}' & =\frac{1}{2}\sqrt{1-2p},
\end{align*}
$(s_{4}',t_{4}')=-(s_{2}',t_{2}')$, and $(s_{5}',t_{5}')=-(s_{3}',t_{3}')$;
for $\gamma=p+\frac{1}{4}$, 
\begin{align*}
s_{1}'' & =t_{1}''=0.
\end{align*}

We now exclude the solutions $(s_{2},t_{2})$ and $(s_{4},t_{4})$
by showing that they are not in the interior of $\mathcal{R}$ when
they are real. To this end, we only need to focus on $(s_{2},t_{2})$
since $(s_{4},t_{4})=-(s_{2},t_{2})$. We prove $(s_{2},t_{2})$ to
be not in the interior of $\mathcal{R}$ by contradiction. We suppose
that $(s_{2},t_{2})$ is real and in the interior of $\mathcal{R}$.
Then, it must hold that $p<\gamma^{2}\le\gamma$ and $2\sqrt{\gamma^{2}-p}>\left|2\gamma-1\right|$,
i.e., $\gamma>p+\frac{1}{4}$. On the other hand, the condition $t^{2}<\gamma^{2}-p$
implies that 
\[
\frac{(\gamma-p)\sqrt{\gamma^{2}-p}}{2\sqrt{\gamma^{2}-p}-\left|2\gamma-1\right|}<\gamma^{2}-p,
\]
which is equivalent to 
\[
2\gamma-1<p<\gamma(2\gamma-1).
\]
Note that $\gamma\neq\frac{1}{2}$. So, $\gamma>1.$ However, this
contradicts with $\gamma\le1$.

We next show that the solutions $(s_{3},t_{3})$ and $(s_{5},t_{5})$
are not in the interior of $\mathcal{R}$ when they are real. To this
end, we only need to focus on $(s_{3},t_{3})$. We suppose that $(s_{3},t_{3})$
is real and in the interior of $\mathcal{R}$. Then, the condition
$t^{2}<\gamma^{2}-p$ implies that $p<\gamma^{2}\le\gamma$ and $0<p<2\gamma-1$.
So, $\gamma>\frac{1}{2}$. On the other hand, the condition $p+s<\gamma+t$
implies that 
\[
\left(2\gamma-1+2\sqrt{\gamma^{2}-p}\right)t_{3}<\gamma-p,
\]
which is equivalent to that $p>2\gamma-1$. This contradicts with
$0<p<2\gamma-1$.

We next show that for $\gamma=\frac{1}{2}$, the solutions $(s_{2}',t_{2}')$,
$(s_{3}',t_{3}')$, $(s_{4}',t_{4}')$, and $(s_{5}',t_{5}')$ are
not in the interior of $\mathcal{R}$. This statement is obvious since
it contradicts with the conditions $t^{2}<\gamma^{2}-p$ and $p\ge0$.

Summarizing all the points, we conclude that $(s,t)=(0,0)$ is the
unique solution (i.e., the unique stationary point) in the interior
of $\mathcal{R}$. At this point, $\varphi(0,0)=0$, i.e., the equality
in \eqref{eq:-58} holds.

We next consider the boundary points of $\mathcal{R}$. We first consider
the boundary points on $t^{2}=\gamma^{2}-p$, i.e., $p=\gamma_{1}\gamma_{2}$.
Statement 1 immediately implies $\varphi(s,t)\ge0$ for this case.
We then consider other boundary points. For this case, $p<\gamma^{2}$.
If $\max\{0,2\gamma-1\}<p<\gamma^{2}$, then by perturbation analysis,
these boundary points cannot be the minimum points. If $p=0$, then
$s=0$, i.e., $p_{1}=p_{2}=0$. By the convexity of $x\mapsto x\log x$,
it is easy to see that the expression in \eqref{eq:-13} is nonnegative,
i.e., $\varphi(s,t)\ge0$, for this case. If $p=2\gamma-1$, then
$\varphi(s,t)\ge0$ follows similarly.

\section{Proof of Lemma \ref{lem:convexity} }

Define 
\begin{align*}
\chi(t,\kappa) & :=-2H\left(\frac{1}{2}+\sqrt{t}\right)+\Big(-(c+\sqrt{t})\log(c+\sqrt{t})\\
 & \qquad-2(\frac{1}{2}-c)\log(\frac{1}{2}-c)-(c-\sqrt{t})\log(c-\sqrt{t})-c\log\kappa\Big).
\end{align*}
Recall that 
\begin{align*}
c & =\frac{\sqrt{\kappa+4\kappa(\kappa-1)t}-1}{2\left(\kappa-1\right)}.
\end{align*}
It is known that $t\mapsto H\left(\frac{1}{2}-\sqrt{t}\right)$ is
concave and nondecreasing \cite{WynerCI}. So, to show that $\chi_{s}$
is convex, it suffices to prove that given any $\kappa\ge1$, the
function $t\mapsto\chi(t,\kappa)$ is convex and nondecreasing.

We compute the first and second derivatives of $t\mapsto\chi(t,\kappa)$
as follows: 
\begin{align*}
\partial_{t}\chi(t,\kappa) & =\frac{1}{2\sqrt{t}}\log\left[\left(\frac{\frac{1}{2}+\sqrt{t}}{\frac{1}{2}-\sqrt{t}}\right)^{2}/\frac{c+\sqrt{t}}{c-\sqrt{t}}\right]\\
 & =\frac{1}{2\sqrt{t}}\log\left[\left(\frac{\frac{1}{2}+\sqrt{t}}{\frac{1}{2}-\sqrt{t}}\right)^{2}/\frac{\sqrt{\kappa+4\kappa(\kappa-1)t}-1+2\left(\kappa-1\right)\sqrt{t}}{\sqrt{\kappa+4\kappa(\kappa-1)t}-1-2\left(\kappa-1\right)\sqrt{t}}\right]
\end{align*}
and 
\begin{align*}
\partial_{t}^{2}\chi(t,\kappa) & =\frac{1}{4t^{3/2}}\Big[-\frac{4\sqrt{t}\left(4(\kappa-1)t-\sqrt{4(\kappa-1)\kappa t+\kappa}+1\right)}{(4t-1)(4(\kappa-1)t+1)}\\
 & \qquad+\log\left(\frac{2(\kappa-1)\sqrt{t}+\sqrt{4(\kappa-1)\kappa t+\kappa}-1}{-2(\kappa-1)\sqrt{t}+\sqrt{4(\kappa-1)\kappa t+\kappa}-1}\right)-2\log\frac{\frac{1}{2}+\sqrt{t}}{\frac{1}{2}-\sqrt{t}}\Big].
\end{align*}
Further, we compute 
\begin{align*}
\partial_{\kappa}\partial_{t}^{2}\chi(t,\kappa) & =\frac{2(\kappa-1)\kappa}{(4(\kappa-1)\kappa t+\kappa)^{3/2}}\ge0.
\end{align*}
So, $\partial_{t}^{2}\chi(t,\kappa)$ is nondecreasing in $\kappa$.
We have $\partial_{t}^{2}\chi(t,\kappa)\ge\lim_{\kappa\downarrow1}\partial_{t}^{2}\chi(t,\kappa)=0$.
So, given any $\kappa\ge1$, $\chi(t,\kappa)$ is convex in $t\in[0,1/4]$.
Furthermore, $\partial_{t}\chi(t,\kappa)\ge\lim_{t\downarrow0}\partial_{t}\chi(t,\kappa)=0$.
So, given any $\kappa\ge1$, $\chi(t,\kappa)$ is nondecreasing in
$t\in[0,1/4]$.

\section{Proof of Proposition \ref{prop:upperbound} }

First, observe that 
\begin{align}
\Gamma_{\alpha}^{\mathrm{UB}}(\pi_{XY}) & =\underset{Q_{XY}}{\sup}\varphi(Q_{XY}),\label{eq:-2}
\end{align}
where 
\begin{align*}
\varphi(Q_{XY}) & :=\underset{\substack{Q_{W|XY}:\\
I_{Q}(X;Y|W)\le(1-1/\alpha)D(Q_{XY}\|\pi_{XY})
}
}{\inf}I_{Q}(XY;W)\\
 & =\underset{\substack{Q_{W|XY},P_{W},P_{X|W},P_{Y|W}:\\
D(Q_{WXY}\|P_{W}P_{X|W}P_{Y|W})\le(1-1/\alpha)D(Q_{XY}\|\pi_{XY})
}
}{\inf}D(Q_{W|XY}\|P_{W}|Q_{XY}).
\end{align*}

We choose $P_{W}P_{X|W}P_{Y|W}$ as the joint distribution of $W\sim\mathrm{Bern}(\frac{1}{2})$,
$X=W\oplus U$, and $Y=W\oplus V$ with $U\sim\mathrm{Bern}(p)$ and
$V\sim\mathrm{Bern}(p)$, where $p\in[0,1]$ and $W,U,V$ are independent.
We choose $Q_{W|XY}$ as the conditional distribution induced by 
\begin{equation}
W=\begin{cases}
X\oplus U, & X=Y\\
V, & X\neq Y
\end{cases},\label{eq:-4}
\end{equation}
where $U\sim\mathrm{Bern}(q)$ and $V\sim\mathrm{Bern}(\frac{1}{2})$
with $q\in[0,1]$ are independent of $(X,Y)$. Denote $\mathcal{Q}$
as the set of such distribution tuples $(Q_{W|XY},P_{W},P_{X|W},P_{Y|W})$.

Using this choice, we obtain the following upper bound on $\varphi(Q_{XY})$.
\[
\varphi(Q_{XY})\le\hat{\varphi}(Q_{XY}):=\underset{\substack{(Q_{W|XY},P_{W},P_{X|W},P_{Y|W})\in\mathcal{Q}:\\
D(Q_{WXY}\|P_{W}P_{X|W}P_{Y|W})\le(1-1/\alpha)D(Q_{XY}\|\pi_{XY})
}
}{\inf}D(Q_{W|XY}\|P_{W}|Q_{XY}).
\]
Denote $r=Q_{XY}(0,1)+Q_{XY}(1,0)$, $c_{0}=q\log\frac{\bar{\epsilon}q}{p^{2}}+\bar{q}\log\frac{\bar{\epsilon}\bar{q}}{\bar{p}^{2}}$,
and $c_{1}=\log\frac{\epsilon}{2p\bar{p}}$. By simple algebra, for
\[
(Q_{W|XY},P_{W},P_{X|W},P_{Y|W})\in\mathcal{Q},
\]
it holds that 
\[
D(Q_{W|XY}\|P_{W}|Q_{XY})=\bar{r}\left(1-H(q)\right),
\]
and 
\begin{align*}
 & D(Q_{WXY}\|P_{W}P_{X|W}P_{Y|W})-(1-1/\alpha)D(Q_{XY}\|\pi_{XY})\\
 & =\bar{r}c_{0}+rc_{1}+\frac{1}{\alpha}D(Q_{XY}\|\pi_{XY})\\
 & \le\bar{r}c_{0}+rc_{1}+\frac{1}{\alpha}D(r\|\epsilon).
\end{align*}
So, we will make $\hat{\varphi}(Q_{XY})$ larger by adjusting $Q_{XY}$
to $\mathrm{DSBS}(r)$. Hence, 
\begin{equation}
\Gamma_{\alpha}^{\mathrm{UB}}(\pi_{XY})\le\underset{r\in[0,1]}{\sup}\hat{\varphi}(\mathrm{DSBS}(r)).\label{eq:-3-1}
\end{equation}

We next prove that 
\begin{equation}
\hat{\varphi}(\mathrm{DSBS}(r))=C(r,(1-1/\alpha)D(r\|\epsilon)).\label{eq:-14}
\end{equation}
To this end, we need the following properties of relaxed Wyner's
common information. 
\begin{prop}
\label{prop:The-relaxed-Wyner} The $t$-relaxed Wyner's common information
$C(r,t)$ of $\mathrm{DSBS}(r)$ satisfies the following properties. 
\begin{enumerate}
\item For $r\in[0,1]$ and $t\ge0$, $C(1-r,t)=C(r,t)$. 
\item Given $r$, $C(r,t)$ is nonincreasing in $t\ge0$. 
\item Given $t$, $C(r,t)$ is nonincreasing in $r\in[0,1/2]$. 
\end{enumerate}
\end{prop}
\begin{IEEEproof}
Statements 1 and 2 follow by definition. We now prove Statement 3.
Consider $0\le r_{1}<r_{2}\le1/2$. Let $(X_{1},Y_{1})\sim\mathrm{DSBS}(r_{1})$.
Let $W$ be an optimal random variable attaining $C(r_{1},t)$. Let
$X_{2}=X_{1}\oplus U$ and $Y_{2}=Y_{1}\oplus V,$ where $U,V$ are
Bernoulli random variables independent of $(X_{1},Y_{1},W)$ such
that $(X_{2},Y_{2})\sim\mathrm{DSBS}(r_{2})$. So, the Markov chain
$W\leftrightarrow(X_{1},Y_{1})\leftrightarrow(X_{2},Y_{2})$ holds,
which implies $I(X_{1}Y_{1};W)\ge I(X_{2}Y_{2};W)$. Similarly, the
Markov chains $(W,Y_{1},Y_{2})\leftrightarrow X_{1}\leftrightarrow X_{2}$
and $(W,X_{1},X_{2})\leftrightarrow Y_{1}\leftrightarrow Y_{2}$ hold,
which implies $I(X_{1};Y_{1}|W)=I(X_{1}X_{2};Y_{1}Y_{2}|W)\ge I(X_{2};Y_{2}|W)$.
So, $C(r_{1},t)\ge C(r_{2},t)$. 
\end{IEEEproof}
On one hand, 
\begin{align*}
\hat{\varphi}(\mathrm{DSBS}(r))\ge\varphi(\mathrm{DSBS}(r)) & =C(r,(1-1/\alpha)D(r\|\epsilon)).
\end{align*}
On the other hand, if we choose $(Q_{W|XY},P_{W},P_{X|W},P_{Y|W})\in\mathcal{Q}$
(described around \eqref{eq:-4}) with $q$ satisfying the conditions
in Proposition \ref{prop:For-a-DSBS} and $p=\bar{r}q+\frac{r}{2}$,
then we can see that for a DSBS $Q_{XY}=\mathrm{DSBS}(r)$, 
\begin{align*}
D(Q_{W|XY}\|P_{W}|Q_{XY}) & =\bar{r}\left(1-H(q)\right),\\
D(Q_{WXY}\|P_{W}P_{X|W}P_{Y|W}) & =(1-1/\alpha)D(r\|\epsilon).
\end{align*}
This implies $\hat{\varphi}(\mathrm{DSBS}(r))\le C(r,(1-1/\alpha)D(r\|\epsilon)).$
Combining the two points above yields \eqref{eq:-14}.

From \eqref{eq:-14}, it holds that 
\begin{align*}
\Gamma_{\alpha}^{\mathrm{UB}}(\pi_{XY}) & \le\underset{r\in[0,1]}{\sup}C(r,(1-1/\alpha)D(r\|\epsilon)).
\end{align*}

We can restrict $r\in[0,1/2]$ in the supremization above. This is
because, it holds that $C(1-r,(1-1/\alpha)D(1-r\|\epsilon))\le C(r,(1-1/\alpha)D(r\|\epsilon))$
for $r\in[0,1/2]$, which follows by Statements 1 and 2 in Proposition
\ref{prop:The-relaxed-Wyner} and the fact that $D(1-r\|\epsilon)\ge D(r\|\epsilon)$
for $r,\epsilon\in[0,1/2]$. We can further restrict $r\in[0,\epsilon]$.
This is because, for $r\in(\epsilon,1/2]$, it holds that $C(\epsilon,0)\ge C(\epsilon,(1-1/\alpha)D(r\|\epsilon))\ge C(r,(1-1/\alpha)D(r\|\epsilon))$.
Therefore, 
\begin{align*}
\Gamma_{\alpha}^{\mathrm{UB}}(\pi_{XY}) & \le\underset{r\in[0,\epsilon]}{\sup}C(r,(1-1/\alpha)D(r\|\epsilon)).
\end{align*}

Furthermore, by restricting $Q_{XY}$ to $\mathrm{DSBS}(r)$ with
$r\in[0,\epsilon]$, it holds that 
\begin{equation}
\Gamma_{\alpha}^{\mathrm{UB}}(\pi_{XY})\ge\underset{r\in[0,\epsilon]}{\sup}\varphi(\mathrm{DSBS}(r))=\underset{r\in[0,\epsilon]}{\sup}C(r,(1-1/\alpha)D(r\|\epsilon)).\label{eq:-3-1-2}
\end{equation}

Hence, Proposition \ref{prop:upperbound} holds.

 \bibliographystyle{unsrt}
\bibliography{ref}

\end{document}